\documentclass[times,twocolumn,final]{elsarticle}

\usepackage{medima}
\usepackage{diagbox}
\usepackage{array, makecell}
\usepackage{framed,multirow}
\newcolumntype{P}[1]{>{\centering\arraybackslash}p{#1}}
\usepackage{booktabs}
\usepackage{threeparttable}
\usepackage{amssymb,amsthm,mathtools}
\usepackage{latexsym}
\usepackage{pifont}

\usepackage{url}
\usepackage{xcolor,colortbl}

\usepackage[colorlinks]{hyperref}
\definecolor{newcolor}{rgb}{.8,.349,.1}
\usepackage[nameinlink,capitalize]{cleveref}
\Crefformat{figure}{#2Fig.~#1#3}
\Crefmultiformat{figure}{Figs.~#2#1#3}{ and~#2#1#3}{,~#2#1#3}{ and~#2#1#3}

\definecolor{brats}{RGB}{33,95,154}
\definecolor{brats_ssa}{RGB}{78,149,217}
\definecolor{washu}{RGB}{70,177,225}
\definecolor{acdc}{RGB}{196,78,82}

\definecolor{mypurple}{RGB}{216,110,204}
\definecolor{glioma}{RGB}{33,95,154}
\definecolor{cardiac}{RGB}{196,78,82}

\journal{Medical Image Analysis}

\begin{document}

\verso{Peijie Qiu \textit{et~al.}}

\begin{frontmatter}

\title{QCResUNet: Joint Subject-level and Voxel-level Segmentation Quality Prediction}%

\author[1]{Peijie \snm{Qiu}}
\cortext[cor1]{Corresponding author: 
  }
\author[2]{Satrajit \snm{Chakrabarty}}
\author[3]{Phuc \snm{Nguyen}}
\author[2]{Soumyendu Sekhar \snm{Ghosh}}
\author[1,4]{Aristeidis \snm{Sotiras}\corref{cor1}}
\ead{aristeidis.sotiras@wustl.edu}

\address[1]{Mallinckrodt Institute of Radiology,
 Washington University School of Medicine, St. Louis, MO, USA}
\address[2]{Department of Electrical and Systems Engineering, Washington University in St. Louis, St. Louis, MO, USA}
\address[3]{Department of Biomedical Engineering, University of Cincinnati, Cincinnati, OH, USA}
\address[4]{Institute for Informatics, Data Science \& Biostatistics, Washington University School of Medicine, St. Louis, MO, USA}


\begin{abstract}
Deep learning has made significant strides in automated brain tumor segmentation from magnetic resonance imaging (MRI) scans in recent years. However, the reliability of these tools is hampered by the presence of poor-quality segmentation outliers, particularly in out-of-distribution samples, making their implementation in clinical practice difficult. Therefore, there is a need for quality control (QC) to screen the quality of the segmentation results. Although numerous automatic QC methods have been developed for segmentation quality screening, most were designed for cardiac MRI segmentation, which involves a single modality and a single tissue type. Furthermore, most prior works only provided subject-level predictions of segmentation quality and did not identify erroneous parts segmentation that may require refinement. To address these limitations, we proposed a novel multi-task deep learning architecture, termed QCResUNet, which produces subject-level segmentation-quality measures as well as voxel-level segmentation error maps for each available tissue class. 
To validate the effectiveness of the proposed method, we conducted experiments
on assessing its performance on evaluating the quality of two distinct segmentation tasks. First, we aimed to assess the quality of brain tumor segmentation results. For this task, we performed experiments on one internal (Brain Tumor Segmentation (BraTS) Challenge 2021, $n = 1,251$) and two external datasets (BraTS Challenge 2023 in Sub-Saharan Africa Patient Population (BraTS-SSA), $n = 40$; Washington University School of Medicine (WUSM), $n = 175$). Specifically, we first performed a three-fold cross-validation on the internal dataset using segmentations generated by different methods at various quality levels, followed by an evaluation on the external datasets. Second, we aimed to evaluate the segmentation quality of cardiac Magnetic Resonance Imaging (MRI) data from the Automated Cardiac Diagnosis Challenge (ACDC, $n=100$).
The proposed method achieved high performance in predicting subject-level segmentation-quality metrics and accurately identifying segmentation errors on a voxel basis. This has the potential to be used to guide human-in-the-loop feedback to improve segmentations in clinical settings.

\end{abstract}


\end{frontmatter}

\section{Introduction}

Medical image segmentation plays an indispensable role for accurate diagnosis, monitoring, treatment planning, and population studies of diseases in modern medicine by enabling precise delineation of anatomical structures and pathological regions~\citep{garcia2017review,litjens2017survey}.
In particular, precise segmentation of healthy and abnormal anatomy into multiple classes using magnetic resonance imaging (MRI) is crucial in this process. 
Recently, deep learning-based methods have achieved state-of-the-art performance in automated segmentation tasks including brain tumor~\citep{unet,deepmedic,modunet,nnunet,brats1} and cardiac~\citep{tran2016fully,khened2019fully,zhou2021nnformer} MRI segmentation.
However, deep neural networks are sensitive to data distribution. This makes them prone to reduced performance when applied on out-of-distribution MRI scans due to variations in acquisition protocols, contrast, image quality, etc. Therefore, quality control (QC) is necessary to thoroughly assess segmentations before they are used for clinical purposes or large-scale research studies.
QC tools are required to detect severe segmentation failures on a per-case basis, pinpoint areas needing segmentation refinement at the voxel level, and provide a quality measure for downstream analyses.
Previously proposed QC methods fall into four main categories: uncertainty estimation-based, generative model-based, reverse classification accuracy (RCA)-based, and regression-based.

The first category of automated segmentation QC methods operates on the premise that high uncertainty is reflective of poor quality segmentations~\citep{ng2018estimating,ALBA2018129,sander2020automatic,ng2020estimating,bai2018automated,uncertainty,jungo2020analyzing,mehta2022qu}. 
Accordingly, most studies focused on developing uncertainty measures to aggregate voxel-wise uncertainty as a proxy for segmentation quality measures, such as Dice Similarity Coefficient (DSC). However, most developed proxy measures~\citep{ng2018estimating,uncertainty,jungo2020analyzing} have not demonstrated a strong correlation with DSC. Moreover, non-negligible errors in uncertainty estimation were observed at the voxel level resulting in unreliable subject-level uncertainty aggregation~\citep{jungo2020analyzing}. Although there are also segmentation methods with built-in quality control~\citep{kalkhof2023m3d}, these methods can only be applied to deep learning-based models and thus cannot be used to assess the quality of segmentations obtained by other methods.

The second category of automated QC methods is based on the assumption that there is a relationship between image intensities and tissue labels.
Along that direction, ~\cite{grady2012automatic} proposed a Gaussian mixture model with hand-crafted features (i.e., geometric features, intensity features, gradient features, and ratio features) to characterize the segmentation quality and detect segmentation failure. ~\cite{deepgen} proposed a variational autoencoder (VAE) to learn the latent representation of pairs of images and their ground-truth segmentations. During inference, the encoder is frozen while the decoder is refined for a given image-segmentation pair to produce a surrogate segmentation. The subject-level DSC is then computed between the query segmentation and the surrogate segmentation.
The tuning of the decoder is required for each query image, which may be computationally expensive and time-consuming. Leveraging the advances in the image-to-image translation task, \cite{li2022towards} proposed a generative adversarial network (GAN) to generate an informative reference image conditioned on the query segmentation mask that needs to be assessed. An auxiliary network (i.e., difference investigator) is then trained to predict image-level and pixel-level quality by taking the raw input image and the generated reference image as inputs. These approaches have only been validated in cardiac MRI segmentation QC that involves only a single modality and segmentation with regular shapes. Translating it to the more complex brain tumor segmentation QC scenario is difficult due to the presence of multiple modalities and intratumoral tissue heterogeneities, which are challenging to model. 

\begin{table*}[!t]
    \centering
    \caption{A detailed comparison of recent segmentation QC methods chosen from four different categories, including \colorbox{lime!25}{UE-based}, \colorbox{yellow!25}{generative-model-based}, \colorbox{pink!25}{RCA-based}, and \colorbox{brown!25}{regression-based} methods. We chose recent representative methods from each category. }
    \resizebox{0.95\textwidth}{!}{
    \begin{tabular}{l|c|c|c|c|c|c}
    \toprule
         & \makecell{Tissue-specific\\error localization} & \makecell{Support multiple \\quality measures} & \makecell{Evaluated segmentation task} & Cross validation & External evaluation & Code availability \\
         \midrule
         \rowcolor{lime!25}
          \cite{jungo2020analyzing} & \textit{coarse} & \ding{51} & \textit{brain tumor} & \ding{55} & \ding{55} & \ding{51}\\
          \midrule
          \rowcolor{yellow!25}
           \cite{deepgen} & \ding{55} & \ding{51} & \textit{cardiac} & \ding{55} & \ding{51} & \ding{55}\\
           \rowcolor{yellow!25}
           \cite{li2022towards} & \ding{51} & \ding{51} & \textit{cardiac} & \ding{55} & \ding{51} & \ding{55}\\
           \midrule
        \rowcolor{pink!25}
         \cite{valindria2017reverse} & \ding{55} & \ding{51} & \textit{whole body} & \ding{51} & \ding{55} & \ding{51}  \\
          \rowcolor{pink!25}
         \cite{robinson2019automated} & \ding{55} & \ding{51} & \textit{cardiac} & \ding{55} & \ding{51} & \ding{55}\\
         \rowcolor{brown!25}
         \midrule
          \cite{robinson2018real} & \ding{55} & \ding{51} & \textit{cardiac} & \ding{55} & \ding{55} & \ding{55} \\ 
          \rowcolor{brown!25}
          \cite{fournel2021medical} & \ding{55} & \ding{55} & \textit{brain tumor} & \ding{55} & \ding{55} & \ding{55}\\
          \rowcolor{brown!25}
            \cite{kofler2022deep} & \ding{55} & \ding{55} & \textit{brain tumor} & \ding{55} & \ding{51} & \ding{55}\\
            \rowcolor{brown!25}
           \cite{qiu2023qcresunet} & \ding{55} & \ding{51} & \textit{brain tumor} & \ding{55} & \ding{51} & \ding{51} \\
           \rowcolor{brown!25}
           \textbf{Ours} & \ding{51} & \ding{51} & \textit{brain tumor \& cardiac} & \ding{51} & \ding{51} & \ding{51}\\
         \bottomrule
    \end{tabular}}
    \label{tab:comparison}
\end{table*}

The third category of the automated segmentation QC methods is built upon the reverse classification accuracy (RCA) framework~\citep{valindria2017reverse}. This was initially designed for whole-body multi-organ MRI segmentation QC and was later applied to cardiac MRI datasets~\citep{robinson2019automated}. 
The RCA framework involves: (i) choosing a reference dataset with known ground-truth segmentation, (ii) training a segmenter using a query image-segmentation pair, (iii)  using the trained segmenter to segment the images in the reference dataset, and (iv) estimating DSC as the maximum DSC achieved by the trained segmenter in the reference dataset.
Although effective for whole-body multi-organ MRI segmentation and cardiac MRI segmentation QC, RCA's reliance on a representative reference dataset and image registration poses challenges in brain tumors. The large variability in brain tumors makes the reference dataset hard to be representative. Different from whole-body multi-organ and cardiac segmentation, which have consistent shapes and appearances, the heterogeneous appearance and phenotypes of brain tumors complicate the establishment of correspondences for the tumorous areas between different subjects. Lastly, despite predicting subject-level DSC, the RCA framework lacks segmentation error localization at the voxel level.

The fourth category of the segmentation QC methods is regression-based methods that directly predict the subject-level DSC.  
Early attempts employed a Support Vector Machine regression in combination with hand-crafted features to detect cardiac MRI segmentation failures~\citep{handfeature,ALBA2018129}. Instead of using hand-crafted features, ~\cite{robinson2018real} proposed a convolutional neural network (CNN) regressor to automatically extract features from a large cardiac MRI segmentation dataset to predict DSC. Besides predicting DSC, \cite{kofler2022deep} proposed a holistic rating to approximate how expert neuroradiologists classify high-quality and poor-quality segmentations. However, manually deriving holistic ratings by clinical experts is laborious and prone to inter-rater variability, hindering its applicability in large-scale datasets. There are also multi-dimensional regression QC approaches~\citep[see e.g.,][]{fournel2021medical} that estimate segmentation quality measures for individual 2D slices and combine them to produce a 3D prediction.  However, this method is specifically designed to predict the DSC, as its formulation relies on combining 2D DSC values into their 3D counterpart.
Additionally, despite the fact that satisfactory performance was achieved in predicting subject-level segmentation metrics by these regression-based methods, voxel-level localization of segmentation error is unavailable. 

Although numerous efforts have been devoted to automated segmentation QC, the existing segmentation QC methods are still limited in many aspects (see a detailed comparison in~\Cref{tab:comparison}). 
First, most studies have focused on evaluating on a single segmentation task, e.g., whole-body multi-organ MRI segmentation and cardiac MRI segmentation QC.  Additionally, most of these approaches have not assessed out-of-distribution generalization on external datasets. Second, there has been limited research on segmentation QC specifically designed for brain tumor MRI segmentation QC. The heterogeneous and complex appearances of brain tumors, which vary in locations, sizes, and shapes, contribute to the increased challenge of QC in brain tumor MRI segmentation. Third, most prior QC approaches have focused on predicting subject-level DSC~\citep{jungo2020analyzing, robinson2018real, robinson2019automated, valindria2017reverse, fournel2021medical,li2022towards} and do not consider or support the quality of the segmentation contour. However, both DSC and contour-based metrics (e.g., normalized surface dice (NSD)) are essential for a comprehensively assessment of the segmentation quality~\citep{maier2024metrics}.
Fourth, it is crucial to address reliable voxel-level tissue-specific segmentation error localization. This localization is vital not only for auditing purposes but also for radiologists to prioritize cases that require manual refinement. By incorporating this component into quality control measures, it can enhance the accuracy and effectiveness of the overall segmentation process. However, this aspect has been largely overlooked in previous studies. One exception is the work proposed by~\cite{li2022towards}. However, this approach was only validated on cardiac segmentation QC with limited datasets. In addition, 
this method can only provide a binary segmentation error mask and is unable to identify voxel-level segmentation errors for different tissue classes, which limits its applicability in clinical practice.

\begin{figure*}[!t]
    \centering
    \includegraphics[width=\textwidth]{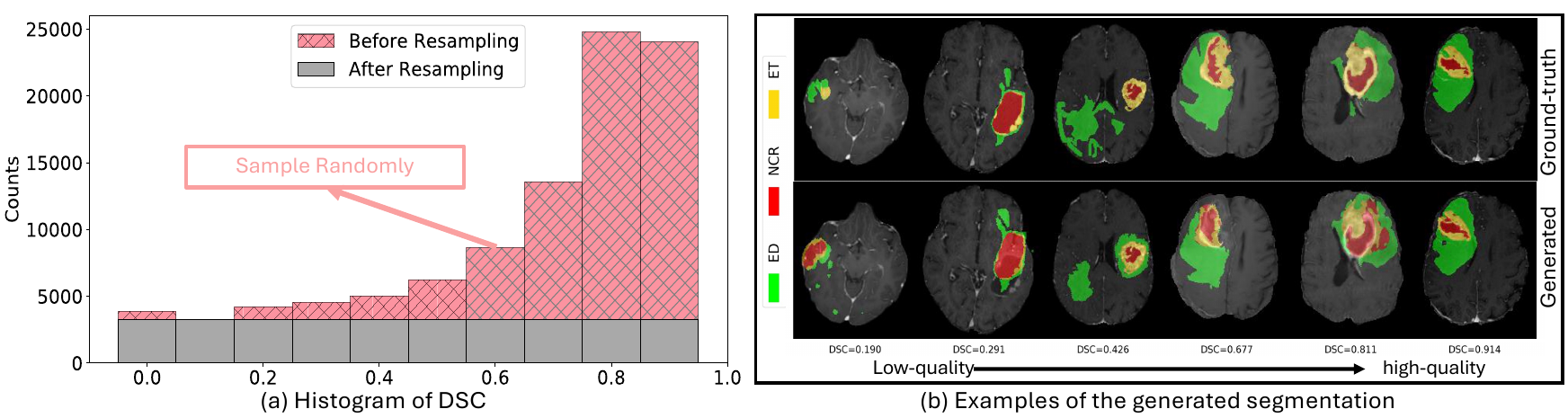}
    \caption{Data generation for the brain tumor segmentation QC task: (a) The histogram of the DSC distribution in the generated BraTS training dataset before and after applying the resampling strategy. (b) Visual examples of the generated segmentation dataset ranging from low-quality to high-quality.}
    \label{fig:dataset}
\end{figure*}

To address these limitations, we proposed a novel deep learning model, termed QCResUNet, to jointly predict segmentation-quality metrics at the subject level and localize segmentation errors across different tissue classes at the voxel level. This study builds upon our previous preliminary work~\citep{qiu2023qcresunet}, but with a focus on developing a more comprehensive and applicable segmentation QC method.
First, we extended the previous work to predict more comprehensive subject-level segmentation measures (DSC and NSD) as well as a collection of binary segmentation error maps, each corresponding to a different tissue class, to provide a more comprehensive evaluation of segmentation quality. To achieve this goal, we further proposed an attention-based segmentation error map aggregation mechanism to better delineate segmentation errors for different tissue classes. Second, we conducted rigorous evaluations to validate the generalizability of the proposed QCResUNet by performing three-fold cross-validation on the internal dataset and evaluating the proposed method on out-of-distribution datasets with MRI scans of varying image quality and segmentations produced. Third, we further examined the generalizability of the proposed QCResUNet by evaluating its ability to assess the segmentation quality for cardiac MRI.
Fourth, we extensively evaluated the performance of the proposed method in comparison with several state-of-the-art segmentation QC methods, including RCA-based~\citep{valindria2017reverse} and uncertainty-based methods~\citep{jungo2020analyzing}.
Fifth, we performed an in-depth explainability analysis to elucidate the reasons behind the proposed method's superior performance compared to other competing approaches.


The main contributions of this work are fivefold: 
\begin{enumerate}
    \item We proposed a multi-task learning framework named QCResUNet  
    to simultaneously predict the DSC and NSD at the subject level and localize segmentation errors at the voxel level.
    \item We proposed an attention-based mechanism to aggregate the segmentation error map that better handles voxel-level segmentation error prediction for different tissue classes.
    \item We extensively evaluated the performance of the proposed model using both internal and external testing for the brain tumor segmentation QC task. Internal training, validation, and testing were performed through a three-fold cross-validation on 1251 cases from Brain Tumor Segmentation (BraTS 2021). External testing was performed on independent datasets from Washington University School of Medicine (WUSM) and BraTS Challenge 2023 in Sub-Saharan Africa Patient Population (BraTS-SSA). Our results demonstrated that the proposed model can generalize well on the out-of-distribution cases from different brain tumor datasets that have been segmented by various methods.
    \item The proposed model was also evaluated on the cardiac MRI segmentation QC task, demonstrating its potential for application in a broader range of QC tasks beyond brain tumor segmentation QC.
    \item To promote reproducibility and foster further research in the community, we have released our code at \url{https://github.com/sotiraslab/QCResUNet}.
\end{enumerate}

\section{Materials and methods}

\subsection{Dataset}
In this study, we used three datasets for evaluation of the QC performance on brain tumor segmentation. First, we used pre-operative multimodal MRI scans with gliomas of all grades (WHO Central Nervous System grades 2-4) grades from the BraTS 2021 challenge training dataset ($n = 1251$). The BraTS dataset is a heterogeneous dataset consisting of cases from 23 different sites with various levels of quality and protocols. The BraTS dataset was used for training, validation, and internal testing.
Additionally, we used two datasets that are not included in the BraTS 2021 dataset (i.e., the BraTS-SSA dataset and the WUSM dataset) for external testing, allowing for an unbiased assessment of the generalizability of the proposed method. 
The BraTS-SSA dataset ($n = 40$)~\citep{adewole2023brain} is an extension of the original BraTS 2021 dataset with patients from Sub-Saharan Africa, which includes lower-quality MRI scans (e.g., poor image contrast and resolution) as well as unique characteristics of gliomas (i.e., suspected higher rates of gliomatosis cerebri). 
The WUSM dataset ($n = 175$) was obtained from the retrospective health records of the Washington University School of Medicine (WUSM), with a waiver of consent, in accordance with the Health Insurance Portability and Accountability Act, as approved by the Institutional Review Board (IRB) of WUSM (\texttt{IRB no.\ PA18-1113}). Each subject in all datasets comprised four modalities viz. pre-contrast T1-weighted (T1), T2-weighted (T2), post-contrast T1-contrast (T1c), and Fluid attenuated inversion recovery (FLAIR). 
In addition, multi-class tumor segmentation masks annotated by experts were also available. Segmentation masks delineated enhancing tumor (ET), non-enhancing tumor core (NCR), and edema (ED) classes. Following standard BraTS procedures, we combined the binary ET, NCR and ED segmentation masks to delineate the whole tumor (WT), tumor core (TC), and enhancing tumor. The WT mask consists of all tumor tissue classes (i.e., ET, NCR, and ED), while the TC mask comprises ET and NCR  tissue classes. 

Scans from the BraTS training and BraTS-SSA datasets were already registered to the SRI24 anatomical atlas~\citep{rohlfing2010sri24}, resampled to 1-mm$^{3}$ isotropic resolution and skull-stripped. For consistency, raw MRI scans from WUSM were pre-processed following the same protocol using the Integrative Imaging Informatics for Cancer Research: Workflow Automation for Neuro-oncology~\citep{chakrabarty2022integrative} framework. 
Subsequently, we z-scored all the skull-stripped scans in the BraTS datasets and the WUSM dataset on a per-scan basis. Finally, scans from the entire dataset were cropped to exclude background regions and then were zero-padded to a common dimension of $160 \times 192 \times 160$ using the nnUNet preprocessing pipeline~\citep{nnunet}. 

In evaluating the QC performance on the cardiac segmentation task, we used the Automated Cardiac Diagnosis Challenge (ACDC) dataset~\citep{bernard2018deep}, which consists of 100 subjects (200 volumes). Each volume in the ACDC dataset is associated with a multi-class segmentation mask delineating the left ventricle (LV), myocardium (Myo), and right ventricle (RV). Lastly, each volume was cropped and zero-padded to a common dimension of $16 \times 16 \times 160$ using the nnUNet pipeline~\citep{nnunet}. 

\subsection{Segmentation Dataset Generation}\label{sec:data_gen}
\label{sec:data_gen}
The majority of previous regression-based methods were trained and validated on datasets limited by both sample size and heterogeneity. To address this limitation, we constructed an extensive dataset consisting of segmentation results of diverse quality levels. This enabled us to capture the variability of segmentations and obtain a reliable performance assessment of the proposed method on a wide range of segmentations. For this purpose, we adopted a four-step approach to produce segmentation results using different methods with various combinations of imaging modalities as input. 

As far as the brain tumor segmentation QC task is concerned, we first employed both CNN-based (i.e., nnUNet~\citep{nnunet}) and Transformer-based (i.e., nnFormer~\citep{zhou2021nnformer}) segmentation UNets and trained them independently seven times by taking different modalities as input (namely T1-only, T1c-only, T2-only, FLAIR-only, T1-FLAIR, T1c-T2, and all four modalities). Since each modality is sensitive to certain tissue types (e.g., T1c is effective in detecting enhancing tumors and FLAIR is advantageous in identifying edema), this approach enabled us to produce segmentations with varying levels of quality. 
We selected these seven different combinations, as these generated sufficiently diverse segmentation samples. Empirical studies~\citep[see e.g.,][]{chakrabarty2022integrative} have shown that including three input modalities does not significantly alter performance compared to only using two. Additionally, when including either T1c and FLAIR, the segmentation model can perform quite well, with the model that takes as input T1c and FLAIR performing as well as the model using all modalities. Based on these observations, we selected modality combinations that maximize segmentation quality diversity while discarding combinations that do not contribute additional variation, thereby reducing unnecessary computational burden. 

Second, we sampled the segmentation along the training routines at various iterations to generate segmentation results from both fully trained and inadequately trained models.  
For this purpose, a small learning rate ($1~\times~10^{-6}$)  was used during training to slower their convergence, allowing the segmentation to improve gradually from low quality to high quality.
Third, we employed a data augmentation strategy based on image transformations to produce diverse segmentation results, termed SegGen. A series of image transformations, including rotation, translation, scaling, and elastic deformation, were randomly applied to each ground-truth segmentation three times with a probability of $0.5$ (see \Cref{tab:segGen}). 
Fourth, apart from the segmentations produced by the aforementioned three methods, we also created out-of-sample segmentation samples from a different segmentation method to assess the generalizability of our proposed model in the testing phase. To accomplish this, we employed the DeepMedic framework~\citep{deepmedic} to train seven segmentation models using the same diverse input modalities as we did in training the nnUNet and nnFormer. After applying the aforementioned four steps, we were able to create a set of erroneous segmentations for the cases included in the internal and external datasets.  The erroneous segmentations along with the ground truth segmentations were subsequently used to estimate the DSC, NSD, and the ground truth segmentation error map (SEM) that were used to train the proposed network. The SEM was computed as the difference between a query segmentation (i.e., one of the erroneous segmentations we produced) and the corresponding ground truth.
Examples of the generated segmentations can be found in~\Cref{fig:dataset}(b).

For the cardiac segmentation task, we employed a procedure similar to the one used in generating segmentation for the brain tumor task. However, unlike brain tumor MRI, cardiac MRI involves only a single modality. Consequently, we trained one nnUNet model and one nnFormer model, sampling the segmentation along the training routines at various iterations. We also generated segmentation results using SegGen during model training. However, due to the technical difficulty of training DeepMedic on the ACDC dataset (i.e., one axis in the ACDC dataset has a significantly smaller input size), we could not include any segmentation produced by DeepMedic for the cardiac segmentation QC task. 

\begin{table}[!t]
    \centering
    \caption{The parameters of image transformations used in SegGen.}
    \vspace{0.3cm}
    \begin{tabular}{c|cc}
    \toprule
         Transformation & Parameters & Probability  \\
         \hline
         Rotation & [$-15^{\circ}$, $15^{\circ}$] & 0.5 \\
         Scaling & scales=[$0.85$, $1.25$] & 0.5 \\
         Translation & moves=[$-20$, $20$] & 0.5 \\
          Deformation &displacements=[-20, 20]  & 0.5 \\
         \bottomrule
    \end{tabular}
    \label{tab:segGen}
\end{table}

\begin{figure*}[!t]
    \centering
    \includegraphics[width=\textwidth]{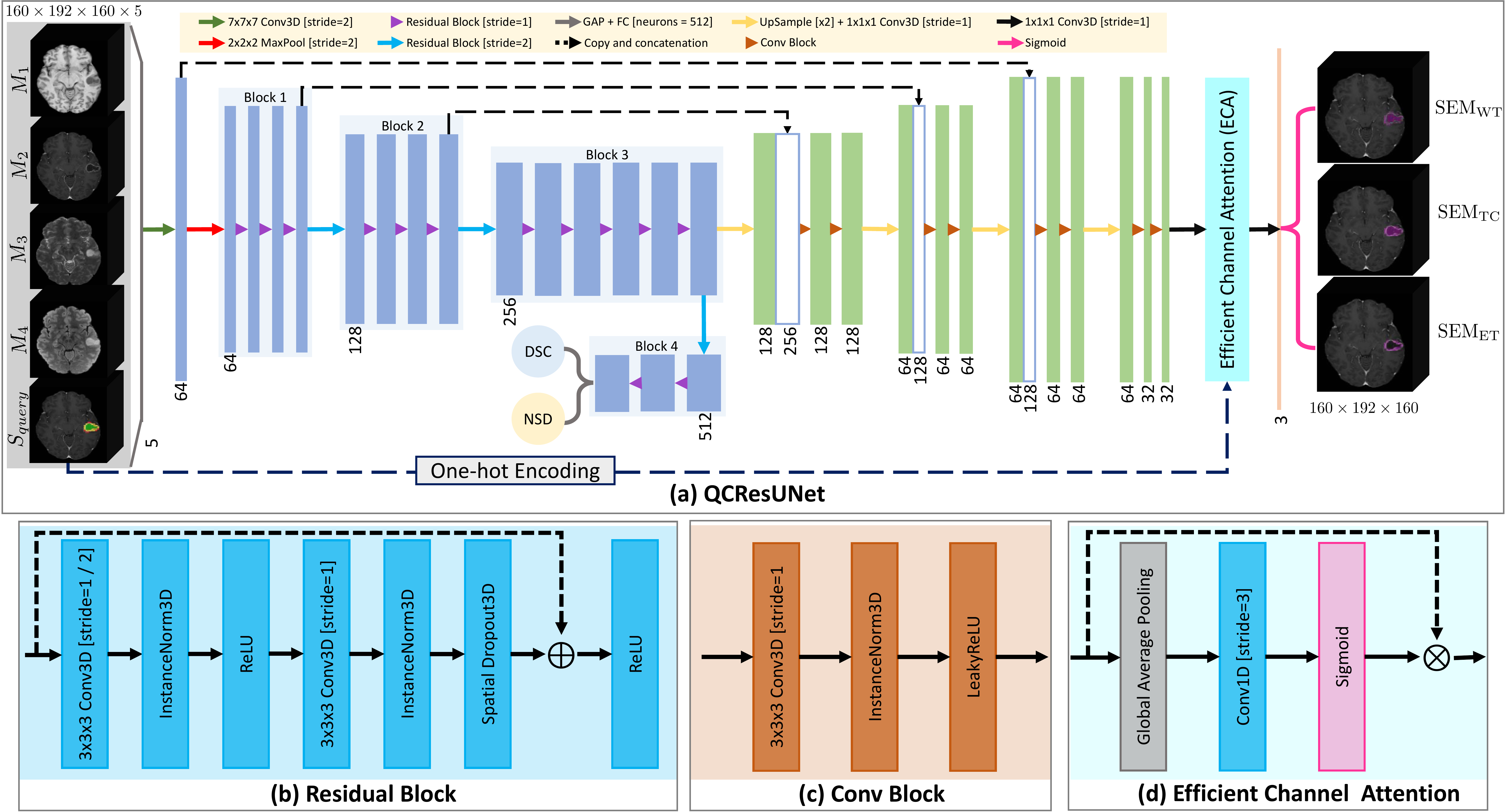}
    \caption{(a) The proposed QCResUNet model is a U-shaped neural network that takes as input $m$ imaging modalities ($\{M_1, M_2, \cdots, M_m \}$) and a multi-class segmentation mask to be evaluated ($S_{query}$). It generates three outputs: the subject-level segmentation-quality metrics DSC ($\text{DSC}_{pred}$) and NSD ($\text{NSD}_{pred}$) as well as a collection of $C$ binary voxel-level segmentation error maps $\{ \text{SEM}_{\text{tissue}_{c}} \}_{c=1}^{C}$, each for each tissue class. In the case of brain tumor segmentation task, which is demonstrated as an example in this figure, QCResUNet takes as input four imaging modalities ($\{M_1, M_2, M_3, M_4 \}$) and a query segmentation $S_{query}$ that delineates WT, TC, and ET. It produces subject-level DSC and NSD, along with three binary SEMs corresponding to the segmentation error masks for WT, TC, and ET tissue classes. Note that the depicted image sizes as well as the number and size of convolution filters are specific to the brain tumor segmentation tasks. 
    Subpanel figures (b) (c), and (d) depict the Residual Block employed in the encoder of QCResUNet, the Convolutional Block used in its decoder, and the Efficient Channel Attention (ECA) used for multiclass SEM aggregation, respectively. The abbreviations in the figure: Conv3D = 3D convolutional layer, GAP = global average pooling, FC = fully connected layer, LeakyReLU = leaky rectified linear unit, and One-hot Encoding = one-hot encodes the multi-class $S_{query}$ to a collection of binary masks.}
    \label{fig:network}
\end{figure*}

Despite our efforts to sample across different convergence stages by adopting a small learning rate, we observed that the deep learning models could successfully segment the majority of the cases. As a consequence, the resulting dataset was distributed unevenly across the different levels of quality, comprising mostly of higher quality segmentation results (see \Cref{fig:dataset}(a)).
Following the resampling strategy in~\cite{robinson2018real}, we randomly selected a subset of samples from each bin of the DSC histogram. This ensured that the number of segmentations from each bin was equal to the lowest count-per-bin value ($n_s$) across the distribution for all the segmentation datasets (refer to \Cref{fig:dataset}(a)). 
For this purpose, we divided DSC values into 10 evenly spanned bins ranging from 0 to 1. We sampled cases from these bins such that we ensured that all bins contained an equal number of cases. For the training set, we did not manually select the DSC samples. Instead, at each training iteration, we randomly sampled segmentations that were evenly distributed across 10 bins to avoid imbalanced DSC values in each training batch. We kept this process stochastic to let the model see as many samples as possible without skewing its exposure to either low DSC or high DSC samples. This means we randomly sampled $n_s$ segmentations for bins that contained more than $n_s$ samples at each training iteration. Conversely, during the evaluation phase, we opted for a deterministic resampling approach, which enabled unbiased evaluations across various quality levels. In order to account for the variability introduced by the resampling process, we implemented a three-fold cross-validation strategy for all our experiments (see~\Cref{sec:exp}).

\subsection{QCResUNet}
\label{sec:qcresunet}
The proposed 3D U-shaped QCResUNet (\Cref{fig:network}(a)) aims to automatically evaluate the quality of a query multi-class segmentation mask by predicting subject-level segmentation quality measures as well as identifying voxel-level segmentation error maps for each tissue class (SEMs). Without loss of generality, QCResUNet takes as input $m$ imaging modalities represented by $\{M_1, M_2, \cdots, M_m\}$ (e.g., $m=4$ the brain tumor segmentation QC case) and the query multi-class segmentation mask ($S_{query}$). The outputs of QCResUNet include two complementary subject-level segmentation quality measures (DSC and NSD) and a collection of binary voxel-level SEMs, each delineating the segmentation error corresponding to a specific tissue class in $S_{query}$. We would like to point out that we followed the standard definition of DSC for calculating the subject-level DSC for multi-class segmentation masks (see~\ref{sec:appendix_dsc}). 


\subsubsection{Network Design}
The U-shaped QCResUNet is designed to perform both regression and segmentation tasks. Therefore, the proposed QCResUNet consists of three parts that are trained end-to-end: (i) a ResNet-34 encoder for DSC and NSD prediction; (ii) a decoder architecture for predicting the multiclass SEM, and (iii) an attention-based SEM aggregation module. 

\noindent \textbf{Encoder:} For the purpose of predicting subject-level DSC, we adopted a ResNet-34~\citep{he2016deep} architecture as part of the encoding path of our network, which can capture semantically rich features that are important for accurately characterizing segmentation quality.
While retaining the main structure of the 2D ResNet-34 in~\cite{he2016deep}, we made the following modifications to account for the 3D nature of the input data. First, all 2D convolutional and pooling layers were replaced by 3D counterparts (see \Cref{fig:network}(b)). Second, we replaced all batch normalization blocks~\citep{ioffe2015batch} with instance normalization blocks~\citep{ulyanov2016instance} to cater to the small batch size during 3D model training. Third, to prevent overfitting, spatial dropout~\citep{tompson2015efficient} was added to each residual block with a probability of 0.3 to randomly zero out channels in the feature map (see \Cref{fig:network}(b)). 

\noindent \textbf{Decoder:} To estimate a segmentation error map with high spatial resolution and accurate localization information, we designed a decoder architecture. The decoder takes the low-resolution contextual features of segmentation quality that were extracted by the encoder and transfers them to a high-resolution multiclass SEM.
This was accomplished by first upsampling the input feature map by a factor of two using nearest neighbor interpolation, followed by a $1 \times 1 \times 1$ convolutional layer. Next, we concatenated the upsampled feature maps with the corresponding encoder level's features using skip-connections to facilitate information flow from the encoder to the decoder. This was followed by two convolutional blocks that reduced the number of feature maps back to the original value before concatenation. Each convolutional block comprised a $3 \times 3 \times 3$ convolutional layer, followed by an instance normalization layer and a Leaky Rectified Linear Unit (LeakyReLU) activation function~\citep{maas2013rectifier} (see \Cref{fig:network}(c)). 

Importantly, we used the middle-level semantics in the intermediate block (Block 3 in \Cref{fig:network}(a)) of the ResNet encoder as the input to the decoder. The rationale behind this was that the last block (Block 4 in \Cref{fig:network}(a)) of the ResNet encoder contains features that are specific to the DSC prediction task. In contrast, the middle-level features are likely to constitute a more universal semantic representation that characterizes the segmentation error~\citep{ahn2019weakly}. 

\noindent \textbf{Attention-based SEM aggregation:} To better predict the segmentation error map corresponding to different tissue classes (e.g., [$\text{SEM}_{\text{WT}}$, $\text{SEM}_{\text{TC}}$, $\text{SEM}_{\text{ET}}$] in the brain tumor segmentation QC task), we propose an attention-based SEM aggregation mechanism. This is because each tissue class in the query segmentation mask ($S_{query}$) contributes differently to the predicted SEM. Specifically, we leveraged the efficient channel attention (ECA)~\citep{wang2020eca} to model the correlation between the features in the last layer of the decoder and the one-hot encoded query segmentation mask. A $1\times 1 \times 1$ convolution layer was then applied to the output of the ECA module to predict tissue-level SEMs (\Cref{fig:network}(a)). Specifically, the channel attention was computed by feeding the pooled input feature (after the global average pooling layer) to a 1D convolutional layer followed by a Sigmoid activation (\Cref{fig:network}(d)).

\subsubsection{Network configuration}
The overall model configuration used for both the brain tumor and cardiac segmentation QC tasks was identical to the one outlined in~\Cref{fig:network}. The only exception is that we used different downsampling/upsampling rates for the brain tumor and cardiac cases to accommodate differences in their respective input sizes. Specifically, the dowsampling/upsampling rates used for the brain tumor case were [[2, 2, 2], [2, 2, 2], [2, 2, 2], [2, 2, 2]] with each [2, 2, 2] represents the downsampling/upsampling rates for axial, sagittal, and coronal axes at each stage. In contrast, the dowsampling/upsampling rates used for the cardiac case were [[1, 2, 2], [1, 2, 2], [1, 2, 2], [2, 2, 2]]. This is because the first axis in the ACDC dataset is ten times smaller than the other two axes.

\subsubsection{Multi-task learning objective}
The proposed QCResUNet was jointly trained for predicting DSC, NSD, and SEM by optimizing the combination of a mean absolute error (MAE) loss, a DSC loss, and a cross-entropy (CE) loss. 

The DSC and NSD prediction task was trained by minimizing the MAE loss ($\mathcal{L}_{MAE}$):
\begin{equation}
    \mathcal{L}_{\text{MAE}} = \frac{1}{N} \sum_{n=1}^{N} \left\{|\text{DSC}_{gt}^{(n)} - \text{DSC}_{pred}^{(n)}| + |\text{NSD}_{gt}^{(n)} - \text{NSD}_{pred}^{(n)}|\right\},
\nonumber 
\end{equation}
where $N$ is the total number of samples in the training dataset, and $n$ is indexing each sample.
The MAE loss ($\mathcal{L}_{MAE}$) quantifies the dissimilarities between the ground truth DSC/NSD (DSC$_{gt}$/NSD$_{gt}$) and the predicted DSC/NSD (DSC$_{pred}$/NSD$_{pred}$).
Compared to the mean squared error loss commonly used for regression tasks, the MAE loss has been proved to be more sensitive to outliers~\citep{mae}. This enhances the robustness of QCResUNet to outliers. 

The voxel-level segmentation error prediction task was trained by optimizing both the DSC and CE losses. This is because combining CE and DSC loss has become mainstream for training models involving dense voxel/pixel-level predictions~\citep{nnunet,zhou2021nnformer,chen2021transunet,cao2022swin,ma2021loss}.
Though there are many variants of the dice loss, we opted for the one proposed in~\cite{drozdzal2016importance} due to its wide success in a variety of medical imaging segmentation tasks:
\begin{equation}
    \begin{split}
        \mathcal{L}_{\text{DSC}} &= - \frac{1}{V}\sum_{c=1} ^{C}\frac{2 \cdot \sum_{v=1}^{V} \text{SEM}_{gt}^{(c,v)} \cdot  \text{SEM}_{pred}^{(c,v)} }{\sum_{v=1}^{V}\text{SEM}_{gt}^{(c,v)} + \sum_{v=1}^{V}\text{SEM}_{pred}^{(c,v)}},
    \end{split}
\nonumber 
\end{equation}
where $V$ is the total number of voxels in a batch, and $v$ indexes each voxel; $C$ is the total number of tissue classes, and $c$ indexes each class.
Here, SEM$_{gt}^{(c)}$ and SEM$_{pred}^{(c)}$ refer to the ground truth for the $c$ tissue class SEM and its probabilistic prediction generated from the decoder's sigmoid output, respectively.
The CE loss is defined as 
\begin{equation}
    \begin{split}
        \mathcal{L}_{\text{CE}} = -\frac{1}{V}\sum_{v=1}^{V} \frac{1}{C}\sum_{c=1}^{c} \text{SEM}_{gt}^{(c, v)} \log \text{SEM}_{pred}^{(c, v)}
    \end{split}
\nonumber 
\end{equation}
The average of the dice loss and CE loss was performed over the total number of voxels ($V$) present in a batch. 

To balance the loss for the subject-level DSC and NSD prediction task and the voxel-level SEM prediction task, we combined the three losses in a weighted fashion. The final objective function is given as 
\begin{equation}
    \mathcal{L}_{final} = \mathcal{L}_{\text{MAE}} + \lambda \ (\mathcal{L}_{\text{DSC}} + \mathcal{L}_{\text{CE}}),
\nonumber 
\end{equation}
where $\lambda$ is the loss balance parameter.

\section{Experimental Validation}
\label{sec:exp}
We evaluated the proposed method on both brain tumor and cardiac MRI segmentation QC tasks to demonstrate its effectiveness and generalizability across different datasets and tasks.

\subsection{Brain tumor MRI segmentation QC task}
We performed three-fold cross-validation to validate the effectiveness of the proposed method for the brain tumor segmentation QC task. For this purpose, we held out an internal testing set with 251 subjects. In each iteration of the three-fold cross-validation, the remaining BraTS dataset was randomly partitioned into training and validation subsets with 667 and 333 subjects, respectively. The application of the previously described four-step approach and resampling (see Section~\ref{sec:data_gen}) resulted in 98,048 training segmentation samples, 28,800 validation segmentation samples, as well as 39,063 internal testing segmentations (36,144 from nnUNet and nnFormer, 2,919 from DeepMedic). We only used the segmentations generated by the DeepMedic as the testing set. This is because segmentations generated by DeepMedic were not seen during training and validation. The BraTS-SSA and WUSM datasets were used as external testing datasets to validate the generalizability of the proposed method. By employing the previously described four-step approach (see Section~\ref{sec:data_gen}), we extended the original BraTS-SSA, and WUSM datasets to include 6,040 and 26,425 segmentations, respectively. After applying the resampling in Section~\ref{sec:data_gen}, the internal BraTS testing, BraTS-SSA, and WUSM datasets were resampled to include 4,753, 1,204, and 2,693 segmentations, respectively.

We performed hyperparameter tuning to determine two critical hyperparameters in the proposed method: the initial learning rate and the loss balance parameter $\lambda$. Hyperparameter tuning was carried out by performing a Bandit-based Bayesian optimization on the training and validation dataset implemented using the Raytune framework~\citep{liaw2018tune}.
Specifically, we sampled the learning rate from a range of $[1\times 10^{-5}, 2 \times 10^{-4}]$ following a log-uniform distribution. Additionally, the loss balance parameter $\lambda$ was chosen from the following set of values $\{0.1, 1, 2\}$. The optimal balance parameter $\lambda$ for QCResUNet was determined to be 1. 
However, we did not observe a significant difference when varying $\lambda$ from 0.1 to 1 and 2 (see~\Cref{fig:hyper}).
Although the optimal learning rate varied from model to model (refer to \Cref{tab:hyperparam}), it generally fell within the scale of $1 \times 10^{-4}$  as we used a relatively small batch size of 4. We kindly direct the readers to~\ref{appendix:hyper_param} for the detailed learning curve in hyperparameter tuning using RayTune.

Results reported hereafter were obtained using the above optimal hyper-parameters.
For both internal  (BraTS) and external (BraTS-SSA and WUSM) testing sets, we reported the results after performing a model ensemble technique for the proposed method and all baseline methods. This involved averaging the results from three models for DSC and NSD prediction and performing majority voting for the SEM prediction. The mean ($\pm$ standard deviation) of MAEs and DSC$_{\text{SEM}}$ were calculated over all subjects. 

\begin{table}[!t]
    \caption{The optimal hyperparameter settings for different models (i.e., UNet, ResNet-34, ResNet-50, QCResUNet) obtained from performing a random search using Raytune. Please refer to~\ref{appendix:hyper_param} for the details of Raytune hyperparameter tuning.}
    \centering
    \resizebox{0.46\textwidth}{!}{
    \begin{tabular}{l|ccc}
    \toprule
         Method & learning rate  & balance parameter ($\lambda$)  \\
          \hline
        UNet & $0.8 \times 10^{-4}$ & - \\
        ResNet-34 & $1.0 \times 10^{-4}$ & - \\ 
        ResNet-50~\citep{robinson2018real} & $0.9 \times 10^{-4}$ & - \\
        QCResUNet & $2.1 \times 10^{-4}$ & 1.0 \\
    \bottomrule
    \end{tabular}}
    \label{tab:hyperparam}
\end{table}

\subsection{Cardiac MRI segmentation QC task}
In the cardiac MRI segmentation QC task, we use the ACDC dataset for training, validation, and testing. The ACDC dataset was partitioned into training, validation, and testing subsets with a ratio of 7:1:2 by following the partitioning rule outlined in~\citep{zhou2021nnformer,chen2021transunet,cao2022swin} to avoid overlapping of subjects. 
This results in a training set consisting of 140 volumes, a validation set with 20 volumes, and a testing set containing 40 volumes. By employing the same four-step approach as in the brain tumor QC task, the original ACDC training, validation, and testing subsets were extended to include 4,900, 640, and 1,280 segmentations. The testing set was then resampled to include 1,011 segmentations. We used the same hyperparameters from the brain tumor segmentation QC task to train the model for the cardiac segmentation QC task.
Similar to the brain tumor segmentation QC task, we also reported the mean ($\pm$ standard deviation) of MAEs and DSC$_{\text{SEM}}$ over all subjects for the cardiac segmentation QC task.

\begin{table*}[!t]
    \centering
    \caption{The subject-level QC performance on the brain tumor MRI segmentation task was evaluated on the internal BraTS testing dataset, as well as the independent BraTS-SSA, and WUSMs datasets with segmentations generated by nnUNet, nnFormer, and DeepMedic. The best metrics within each column are highlighted in \textbf{bold}.}
    \resizebox{0.95\textwidth}{!}{
     \begin{threeparttable}
    \begin{tabular}{lr|cccc|cc|cc}
    \toprule 
    & \multicolumn{1}{c}{} & \multicolumn{4}{c|}{Internal} & \multicolumn{4}{c}{External}  \\
    \cmidrule(r){3-6} \cmidrule(r){7-10} 
    \multicolumn{2}{c}{\backslashbox[70mm]{\textbf{Method}}{\textbf{Dataset}}} & \multicolumn{4}{c|}{\textcolor{brats}{BraTS [testing]}} & \multicolumn{2}{c}{\textcolor{brats_ssa}{BraTS-SSA}} & \multicolumn{2}{c}{\textcolor{washu}{WUSM}}\\
    \cmidrule(r){3-6} \cmidrule(r){7-8} \cmidrule(r){9-10} 
    & & \multicolumn{2}{c}{nnUNet+nnFormer} & \multicolumn{2}{c|}{DeepMedic} & \multicolumn{2}{c}{nnUNet+nnFormer+DeepMedic} 
     & \multicolumn{2}{c}{nnUNet+nnFormer+DeepMedic} \\
    \cmidrule(r){3-4} \cmidrule(r){5-6} \cmidrule(r){7-8} \cmidrule(r){9-10} 
    & \multicolumn{1}{|c|}{\textbf{Metric}} & $r \uparrow$ & MAE $\downarrow$ & $r \uparrow$ & MAE $\downarrow$ & $r \uparrow$ & MAE $\downarrow$ & $r \uparrow$ & MAE $\downarrow$ \\
    \midrule
    \multirow[t]{2}{*}{RCA~\citep{valindria2017reverse}} & \multicolumn{1}{|c|}{\makecell{NSD \\ DSC}}  & \makecell{ 0.662 \\ 0.624 } & \makecell{0.255 $\pm$ 0.165\\ 0.315 $\pm$ 0.227 } & \makecell{0.631 \\0.660 } & \makecell{0.272 $\pm$ 0.171 \\ 0.329 $\pm$ 0.215 } & \makecell{ 0.690 \\ 0.604 } & \makecell{ 0.260 $\pm$ 0.160 \\ 0.302 $\pm$ 0.227 } &\makecell{0.736 \\ 0.504 } & \makecell{0.242 $\pm$ 0.158 \\ 0.308 $\pm$ 0.246 } \\
    \cline{2-10}
    \multirow[t]{2}{*}{UE-based~\citep{jungo2020analyzing}} & \multicolumn{1}{|c|}{\makecell{NSD \\ DSC}} & \makecell{0.817 \\ 0.784} & \makecell{ 0.102 $\pm$ 0.095 \\ 0.144 $\pm$ 0.112 } & \makecell{0.744 \\ 0.762} & \makecell{ 0.113 $\pm$ 0.083 \\ 0.149 $\pm$ 0.114} & \makecell{0.627 \\ 0.703} & \makecell{0.149 $\pm$ 0.114 \\ 0.165 $\pm$ 0.122} & \makecell{0.591 \\ 0.617} & \makecell{0.170 $\pm$ 0.119 \\ 0.187 $\pm$ 0.141} \\
    \cline{2-10}
     \multirow[t]{2}{*}{Multi-D~\citep{fournel2021medical}} & \multicolumn{1}{|c|}{\makecell{NSD \\ DSC}} & \makecell{ - \\ 0.630} & \makecell{ - \\ 0.262 $\pm$ 0.224 } & \makecell{ - \\ 0.605} & \makecell{ - \\ 0.351 $\pm$ 0.228} & \makecell{ - \\ 0.607} & \makecell{- \\ 0.375 $\pm$ 0.243} & \makecell{ -\\ 0.328} & \makecell{ -\\ 0.375 $\pm$ 0.243} \\
    \cline{2-10}
    \multirow[t]{2}{*}{UNet} & \multicolumn{1}{|c|}{\makecell{NSD \\ DSC}} & \makecell{0.939 \\ 0.944} & \makecell{0.069 $\pm$ 0.053 \\ 0.077 $\pm$ 0.063} & \makecell{0.902 \\ 0.939 } & \makecell{0.087 $\pm$ 0.054 \\ 0.080 $\pm$ 0.069 }  & \makecell{0.946 \\ 0.931} & \makecell{0.061 $\pm$ 0.050 \\ 0.079 $\pm$ 0.074} & \makecell{0.903 \\ 0.885} & \makecell{0.080 $\pm$ 0.073 \\ 0.102 $\pm$ 0.108} \\
    \cline{2-10}
    \multirow[t]{2}{*}{ResNet-34} & \multicolumn{1}{|c|}{\makecell{NSD \\ DSC}} & \makecell{0.944 \\ 0.955} & \makecell{0.070 $\pm$ 0.054 \\ 0.072 $\pm$ 0.057} & \makecell{0.916\\0.954} & \makecell{0.086 $\pm$ 0.065\\ 0.074 $\pm$ 0.059} & \makecell{0.950 \\ 0.943} & \makecell{0.067 $\pm$ 0.052 \\ 0.076 $\pm$ 0.068}& \makecell{0.894 \\ 0.888} & \makecell{0.094 $\pm$ 0.082 \\ 0.103 $\pm$ 0.107}\\
    \cline{2-10}
    \multirow[t]{2}{*}{ResNet-50~\citep{robinson2018real} } & \multicolumn{1}{|c|}{\makecell{NSD \\ DSC}} & \makecell{0.943 \\ 0.960} & \makecell{0.069 $\pm$ 0.052 \\ 0.070 $\pm$ 0.053} & \makecell{0.917\\0.947} & \makecell{0.084 $\pm$ 0.061 \\ 0.074 $\pm$ 0.056} &\makecell{0.946 \\ 0.941} & \makecell{0.066 $\pm$ 0.049 \\ 0.077 $\pm$ 0.068} &\makecell{0.895 \\ 0.895} & \makecell{0.094 $\pm$ 0.082 \\ 0.103 $\pm$ 0.107}
    \\
    \midrule
    \multirow[t]{2}{*}{QCResUNet} & \multicolumn{1}{|c|}{\makecell{NSD \\ DSC}} & \makecell{\textbf{0.958}$^*$ \\ \textbf{0.968}$^*$} & \makecell{\textbf{0.056 $\pm$ 0.043}$^*$ \\ \textbf{0.064 $\pm$ 0.048}$^*$} & \makecell{\textbf{0.937}$^*$ \\ \textbf{0.962}$^*$} & \makecell{\textbf{0.074 $\pm$ 0.050}$^*$ \\ \textbf{0.062 $\pm$ 0.047}$^*$ } & \makecell{\textbf{0.954}$^*$ \\ \textbf{0.964}$^*$}  & \makecell{\textbf{0.057 $\pm$ 0.044}$^*$ \\ \textbf{0.060 $\pm$ 0.049}$^*$} & \makecell{\textbf{0.920}$^*$ \\\textbf{0.912}$^*$}& \makecell{\textbf{0.075 $\pm$ 0.062}$^*$ \\ \textbf{0.087 $\pm$ 0.097}$^*$} \\
    \bottomrule
    \end{tabular}
    \begin{tablenotes}
            \item  $^*$: $P < 0.05$; with a paired t-test to all baseline methods.
    \end{tablenotes}
    \end{threeparttable}
    }
    \label{tab:brain_tumor_subj}
\end{table*}
\begin{table}[!t]
    \centering 
    \caption{The subject-level QC performance on the cardiac MRI segmentation task was evaluated on the internal ACDC testing set with segmentations produced by nnUNer and nnFormer. The best metrics within each column are highlighted in \textbf{bold}.}
    \resizebox{0.45\textwidth}{!}{
    \begin{threeparttable}
    \begin{tabular}{l|cc|cc}
    \toprule 
    \multicolumn{1}{c|}{} & \multicolumn{4}{c}{\textcolor{acdc}{ACDC [testing]}} \\
    \cmidrule(r){2-5}
    \backslashbox[25mm]{\textbf{Method}}{\textbf{Metric}} & \multicolumn{2}{c|}{NSD} & \multicolumn{2}{c}{DSC}  \\
    \cmidrule(r){2-3} \cmidrule(r){4-5}
    & $r \uparrow$ & MAE $\downarrow$ & $r \uparrow$ & MAE $\downarrow$ \\
    \toprule 
    RCA  & 0.760  & 0.185 $\pm$ 0.151  & 0.808 & 0.291 $\pm$ 0.136 \\
    UE-based  & 0.827 & 0.088 $\pm$ 0.064  & 0.859 & 0.085 $\pm$ 0.063  \\
    Multi-D & - & - & 0.884 & 0.166 $\pm$ 0.088\\
    UNet  & 0.884 & 0.071 $\pm$ 0.060  & 0.930 & 0.059 $\pm$ 0.053  \\
    ResNet-34  & 0.891 & 0.073 $\pm$ 0.061  & 0.931 & 0.059 $\pm$ 0.053 \\
    ResNet-50  & 0.883 & 0.076 $\pm$ 0.059  & 0.938 & 0.061 $\pm$ 0.049  \\
    \midrule
    QCResUNet & \textbf{0.914}$^*$ & \textbf{0.070 $\pm$ 0.047}$^*$  & \textbf{0.955}$^*$ & \textbf{0.057 $\pm$ 0.040}$^*$  \\
    \bottomrule
    \end{tabular}
    \begin{tablenotes}
            \item  $^*$: $P < 0.05$; with a paired t-test to all baseline methods.
    \end{tablenotes}
    \end{threeparttable}
    }
    \label{tab:cardiac}
\end{table}

\subsection{Baseline methods}
To evaluate the effectiveness of the proposed model, we compared its performance with five baseline models: (i) RCA method~\citep{rca}; (ii) uncertainty estimation (UE) based method~\citep{jungo2020analyzing}; and three regression-based methods: (iii) a UNet model~\citep{unet}; (iv) a ResNet-34 model~\citep{he2016deep}; and (v) a ResNet-50~\citep{he2016deep} model~\citep{robinson2018real}. 

The selection of the three regression-based methods (UNet, ResNet-34, ResNet-50) as baselines was based on their architectural resemblance to the proposed approach. To ensure a fair comparison, the residual blocks in ResNet-34 and ResNet-50 were identical to those used in the proposed QCResUNet. As the UNet model typically outputs a segmentation mask of the same dimensions as the input, we added an average pooling layer after its final feature map followed by a fully connected layer to enable the prediction of a single DSC value. The RCA and UE-based methods are considered state-of-the-art in the segmentation QC literature.

RCA for brain tumor segmentation QC was implemented using as segmentation method the DeepMedic framework, which was adapted following the protocol described by~\cite{valindria2017reverse}. Specifically, we reduced the number of feature maps in each layer by one third compared to the default setting of DeepMedic. We also reduced the feature maps in the last fully connected layer from 150 to 45. 
We opted for this approach instead of the atlas-based segmentation approach because establishing spatial correspondences between query and reference images in tumorous areas can be challenging due to the high spatial and phenotypic heterogeneity of brain tumors. In contrast, RCA for cardiac MRI segmentation QC was implemented using the atlas-based segmentation method by following the protocols in~\citep{robinson2019automated}. This is because registration correspondences between query and reference are easier to establish in healthy autonomy.
For the selection of the reference data set, we randomly selected 100 samples from the training dataset following a previously well-validated protocol~\citep{robinson2019automated}.

The implementation of the UE-based method followed the protocol in~\cite{jungo2020analyzing}, where the uncertainty map of each generated segmentation sample was obtained by applying the MCDropout~\citep{gal2016dropout} to the trained nnUNet, nnFormer and DeepMedic (see~\ref{sec:appendix_a}). The uncertainty map was then calibrated by evaluating the uncertainty-error (UE) overlap. The UE overlap measures the overlap between the binarized uncertainty map and the corresponding SEM (see~\ref{appendix:UE} for details). We chose the UE calibration proposed by~\cite{jungo2020analyzing} instead of other calibration methods~\citep{wen2020batchensemble,ashukha2020pitfalls,mehta2022qu} because it produces a binary mask by thresholding the uncertainty map. We can directly compare the resulting binary mask to the SEM generated by the proposed QCResUNet. The subject-level DSC was computed by applying a random forest with 102 extracted radiomics features (see~\ref{appendix:subj}). 

\begin{figure*}[!t]
    \centering
    \includegraphics[width=0.8\textwidth]{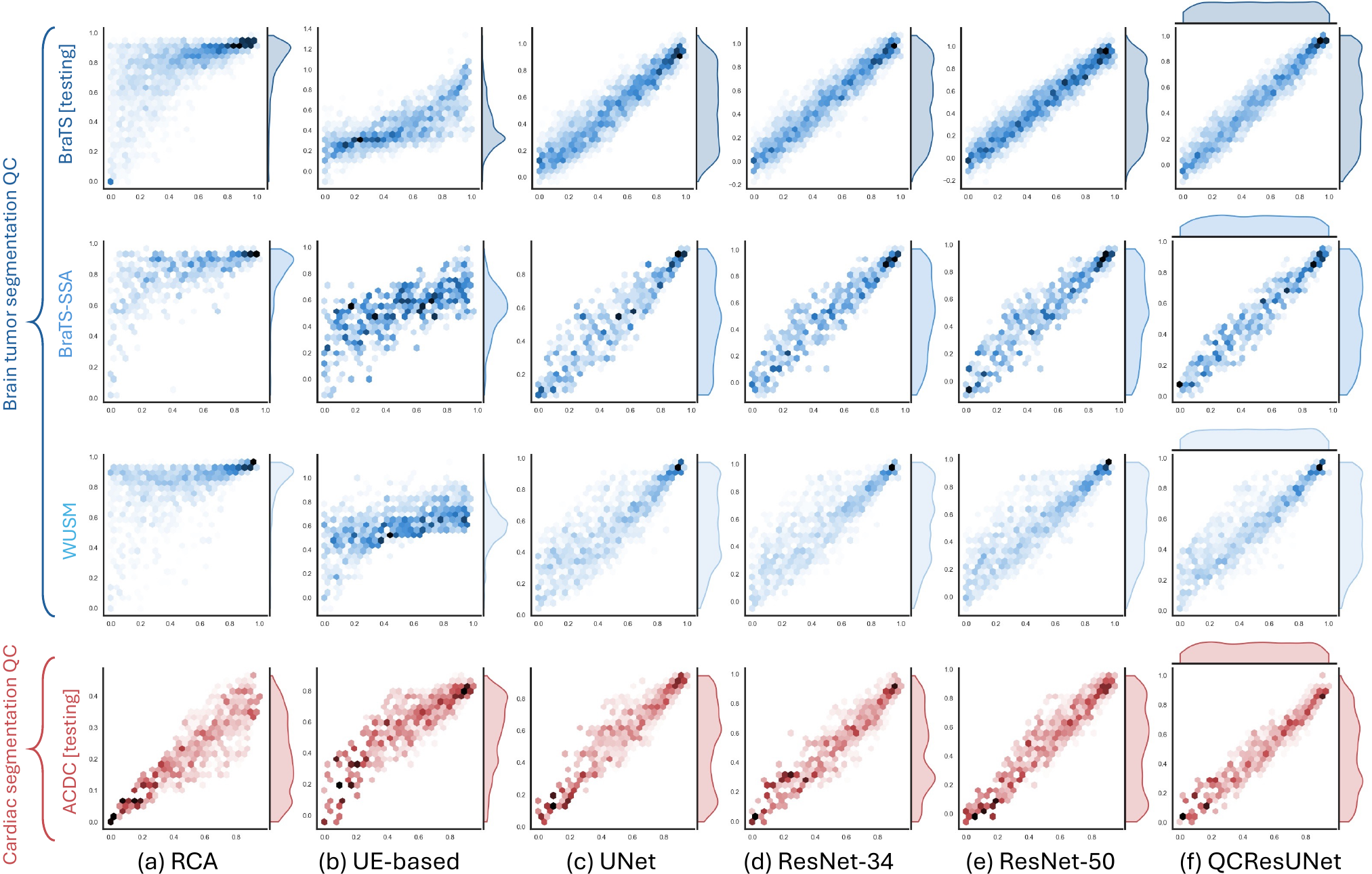}
    \caption{Scatter plots of the ground-truth (x-axis) and the predicted DSC (y-axis) for the proposed method and all the baseline methods in the internal BraTS testing, external BraTS-SSA, WUSM datasets, and ACDC internal testing (rows). Results for (a) RCA; (b) UE-based; (c) UNet; (d) ResNet-34; (e)ResNet-50; and (f) QCResUNet are reported in different columns. The proposed method generalized well to external datasets and consistently showed superior performance compared to all baseline methods. The UE-based method showed the worst performance compared to all other methods.}
    \label{fig:scatter_DSC}
\end{figure*}

\begin{figure*}[!t]
    \centering
    \includegraphics[width=0.8\textwidth]{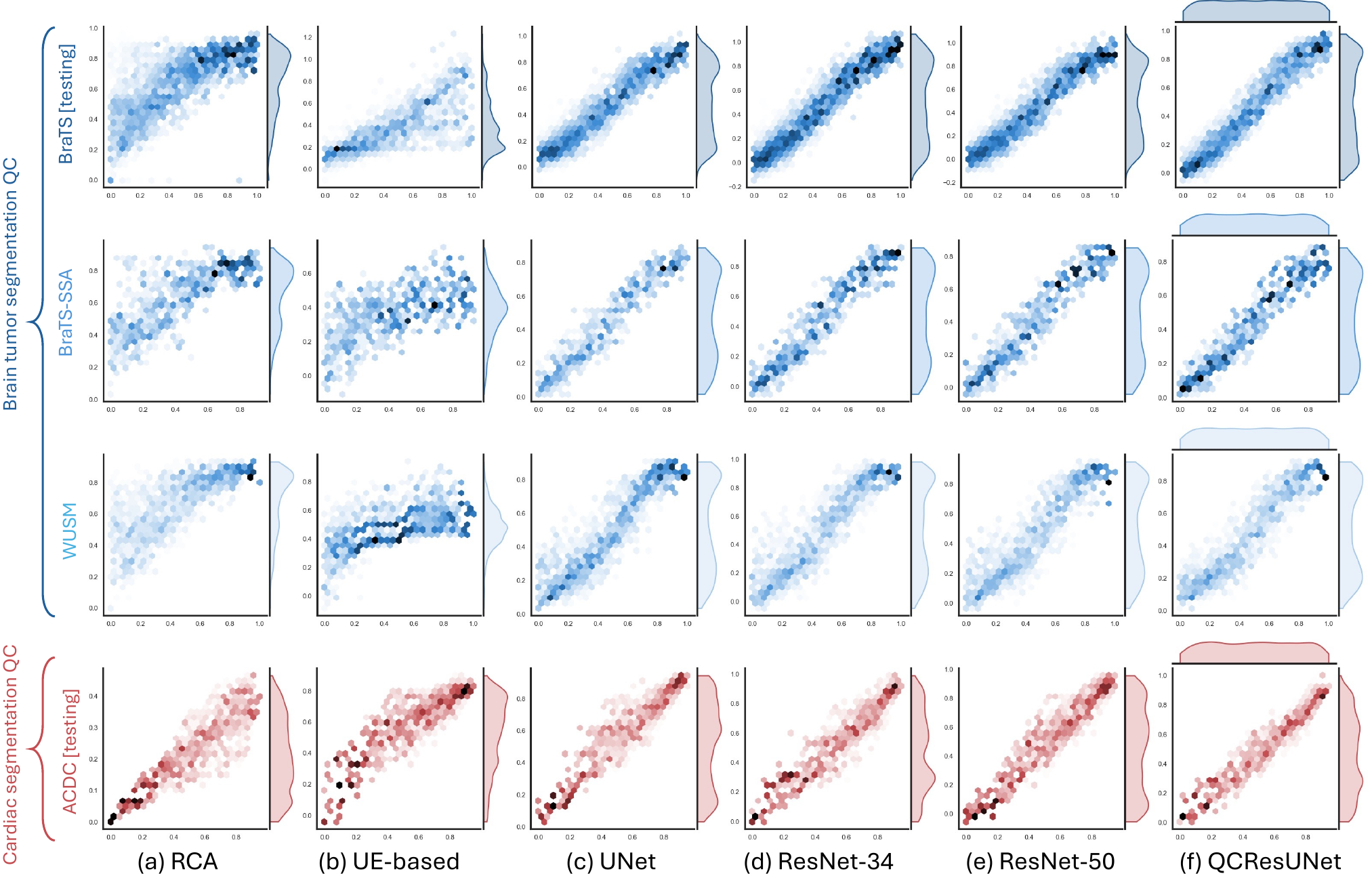}
    \caption{Scatter plots of the ground-truth (x-axis) and the predicted NSD (y-axis) for the proposed method and all the baseline methods in the internal BraTS testing, external BraTS-SSA, WUSM datasets, and ACDC internal testing (rows). Results for (a) RCA; (b) UE-based; (c) UNet; (d) ResNet-34; (e)ResNet-50; and (f) QCResUNet are reported in different columns.  The proposed method generalized well to external datasets and consistently showed superior performance compared to all baseline methods. The UE-based method showed the worst performance compared to all other methods.}
    \label{fig:scatter_NSD}
\end{figure*}

\subsection{Evaluation metrics}
We evaluated the performance of QC segmentation at both the subject level and the voxel level. At the subject level, we evaluated the precision of the prediction of the segmentation quality metric based on the MAE and the Pearson correlation coefficient $r$. MAE measures the amount of error in the prediction. $r$ measures the linear correlation between the predicted DSC/NSD value (DSC$_{pred}$/NSD$_{pred}$) and the ground truth DSC/NSD (DSC$_{gt}$/NSD$_{gt}$). It ranges from $-1$ (strongly uncorrelated) to $+1$ (strongly correlated).

At the voxel level, the performance was evaluated based on the DSC (DSC$_{\text{SEM}}^{(c)}$) between the predicted SEM (SEM$_{pred}^{(c)}$) and ground-truth SEM (SEM$_{gt}^{(c)}$) for each tissue class $c$. Here, we report the average DSC across all $C$ tissue classes.
We performed a paired t-test to compare the DSC predicted by the QCResUNet with those predicted by the corresponding baselines. The resulting p-value ($P$) was reported as a measure to determine if there was a statistically significant difference between the performance of the proposed method and the baseline methods. The significance level was set a priori to $P < 0.05$.

\subsection{Implementation details}
Training procedures were the same for both tasks by using segmentation results were produced by nnUNet, nnFormer, and SegGen. Training was carried out using Adam optimizer~\citep{kingma2014adam} with a batch size of 4 for 100 epochs. Training was started with an optimal initial learning rate determined by the hyperparameter search and was exponentially decayed by a factor of 0.9 for each epoch until it reached $1\times 10^{-6}$. 
We applied an $L_2$ weight decay of $1 \times 10^{-4}$ in all trainable layers. To prevent overfitting, various data augmentations, including random rotation, scaling, mirroring, Gaussian noise, and gamma intensity correction, were applied during training. All model training was performed on NVIDIA Tesla A100 GPUs. By default, we used a mix of single-precision (FP32) and half-precision (FP16) tensors during training to reduce training time and memory consumption.  The proposed method was implemented using \texttt{PyTorch v1.12.1}.

\section{Results}

\subsection{Evaluation of subject-level QC performance}
\subsubsection{Brain tumor MRI segmentation QC task}
The proposed model performed well in the subject-level DSC and NSD prediction across all three brain tumor datasets (\Cref{tab:brain_tumor_subj}). 
Specifically, on the BraTS internal testing set with segmentations generated by nnUNet and nnFormer, the proposed method achieved a small mean MAE of 0.056 and 0.064 for NSD and DSC prediction, respectively. The predicted NSD and DSC also showed a strong correlation with the corresponding ground truth, achieving Pearson r values of 0.958 and 0.968, respectively. Importantly, the proposed method generalized well for segmentation produced by different methods (i.e., DeepMedic), which had not been used during training, achieving an average MAE of 0.074 and 0.062 for NSD and DSC prediction. Similarly, the predicted NSD and DSC showed a strong correlation with their ground truth with Pearson r values of 0.937 and 0.962, respectively.

Critically, our method also generalized well to completely unseen external BraTS-SSA and WUSM datasets.
On the BraTS-SSA dataset, the proposed method achieved an MAE of $0.057$ and $0.060$ for NSD and DSC prediction, respectively. In addition, the Pearson $r$ between the predicted segmentation quality measures and their ground-truth values demonstrated a high correlation (NSD $r = 0.954$; DSC $r = 0.964$).
Despite containing MRI scans with varying image quality and tumor characteristics, the proposed method still generalized well to the BraTS-SSA dataset. However, there was a slight drop in performance on the WUSM dataset with a Pearson r of 0.920 and 0.912 for NSD and DSC predictions, respectively. The MAE for the NSD and DSC prediction on the WUSM dataset was 0.075 and 0.087, respectively.
We conjectured this might be attributed to domain shift due to differences in data acquisition, preprocessing, and the variability of shape and structures in brain tumors. 

Importantly, the proposed method outperformed all baseline methods in NSD and DSC prediction tasks. Compared to the three regression-based QC methods (i.e., UNet, ResNet-34, ResNet-50), QCResUNet improved the second-best method by 0.8\% for NSD prediction and 1.5\% for DSC prediction in terms of Pearson r values on the BraTS internal testing set. On the external BraTS-SSA and WUSM datasets, QCResUNet outperformed the second-best method by an average of 1.3\% and 1.9\% in terms of Pearson r, respectively. A paired t-test confirmed that this improvement was statistically significant compared to all three regression-based methods baseline (\Cref{tab:brain_tumor_subj}). In addition, the proposed method outperformed multi-dimensional regression-based method~\citep{fournel2021medical} in subject-level DSC prediction by an average of 56.3\%, 58.8\%, and 180\% on internal testing, external BraTS-SSA, and external WUSM datasets, respectively (\Cref{tab:brain_tumor_subj}). We hypothesize that the inferior performance of multi-dimensional regression-based methods, compared to other regression-based approaches, is attributed to the aggregation of 2D DSCs across slices without accounting for tumor sizes. As a consequence, each slice contributes equally to the aggregation of 2D DSCs. However, the variability in tumor sizes within a single 3D subject or across different subjects presents a significant challenge in accurately aggregating them for subject-level DSC prediction.
The proposed approach exhibited more evenly distributed DSC prediction errors across different quality levels compared to all the baseline methods (refer to \Cref{fig:scatter_DSC} and \Cref{fig:scatter_NSD}), demonstrating a smaller standard deviation in MAE (see \Cref{tab:brain_tumor_subj}). Furthermore, our QCResUNet offers computational efficiency comparable to ResNet50 and slightly surpasses that of UNet, especially in terms of FLOPs (see more computational benchmarking in~\ref{appendix:computation}).

\begin{table*}[!t]
    \centering 
    \caption{Comparison of voxel-level segmentation QC Performance of the proposed method to baseline methods in terms of DSC$_{\text{SEM}}$. DSC$_{\text{SEM}}$ for the brain tumor MRI segmentation QC task is computed as the average of DSC$_{\text{SEM}}^{\text{WT}}$, DSC$_{\text{SEM}}^{\text{TC}}$, and DSC$_{\text{SEM}}^{\text{ET}}$. While DSC$_{\text{SEM}}$ for cardiac MRI segmentation QC task is computed as the average of DSC$_{\text{SEM}}^{\text{LV}}$, DSC$_{\text{SEM}}^{\text{Myo}}$, and DSC$_{\text{SEM}}^{\text{RV}}$. The best metrics within each column are highlighted in \textbf{bold}.}
    \resizebox{0.9\textwidth}{!}{
    \begin{threeparttable}
    \begin{tabular}{ll|ccc|c}
    \toprule 
    \multicolumn{2}{c|}{} & \multicolumn{3}{c|}{\textcolor{glioma}{Brain tumor segmentation QC}} & \multicolumn{1}{c}{\textcolor{cardiac}{Cardiac segmentation QC}} \\
     \cmidrule(r){3-5} \cmidrule(r){6-6} 
     \multicolumn{2}{c|}{\backslashbox[45mm]{\textbf{Method}}{\textbf{Dataset}}} & \textcolor{brats}{BraTS [testing]}  & \textcolor{brats_ssa}{BraTS-SSA} & \textcolor{washu}{WUSM} & \textcolor{acdc}{ACDC [testing]} \\
    \toprule 
    & UE-based~\citep{jungo2020analyzing} & 0.360 $\pm$ 0.223 & 0.309 $\pm$ 0.106 & 0.283 $\pm$ 0.150 & 0.460 $\pm$ 0.038\\
    \midrule
    \multirow{3}{*}[-0.0cm]{\rotatebox[origin=c]{90}{\makecell{\color{gray!85}\textbf{ablation}}}}  & QCResUNet w/o attn. & 0.633 $\pm$ 0.190 & 0.641 $\pm$ 0.052 & 0.618 $\pm$ 0.100 & 0.613 $\pm$ 0.116 \\
    &QCResUNet w/o regr. & 0.746 $\pm$ 0.164 & 0.685 $\pm$ 0.115 & 0.674 $\pm$ 0.076 & 0.668 $\pm$ 0.091 \\
    &QCResUNet & \textbf{ 0.769 $\pm$ 0.131}$^*$  &  \textbf{0.702 $\pm$ 0.088}$^*$  & \textbf{0.684 $\pm$ 0.073}$^*$ & \textbf{0.703 $\pm$ 0.082}$^*$  \\
    \bottomrule
    \end{tabular}
    \begin{tablenotes}
            \item  $^*$: $P < 0.05$; with a paired t-test to all baseline methods. QCResUNet w/o attn. refers to QCResUNet without the proposed attention-based SEM aggregation. QCResUNet w/o regr. indicates QCResUNet without performing subject-level QC of DSC and NSD. QCResUNet represents the proposed QCResUNet that utilizes attention-based SEM aggregation and performs both subject-level and voxel-level QC prediction.
    \end{tablenotes}
    \end{threeparttable}
    }
    \label{tab:brain_tumor_voxel}
\end{table*}
\begin{figure*}[!t]
    \centering
    \includegraphics[width=1.0\textwidth]{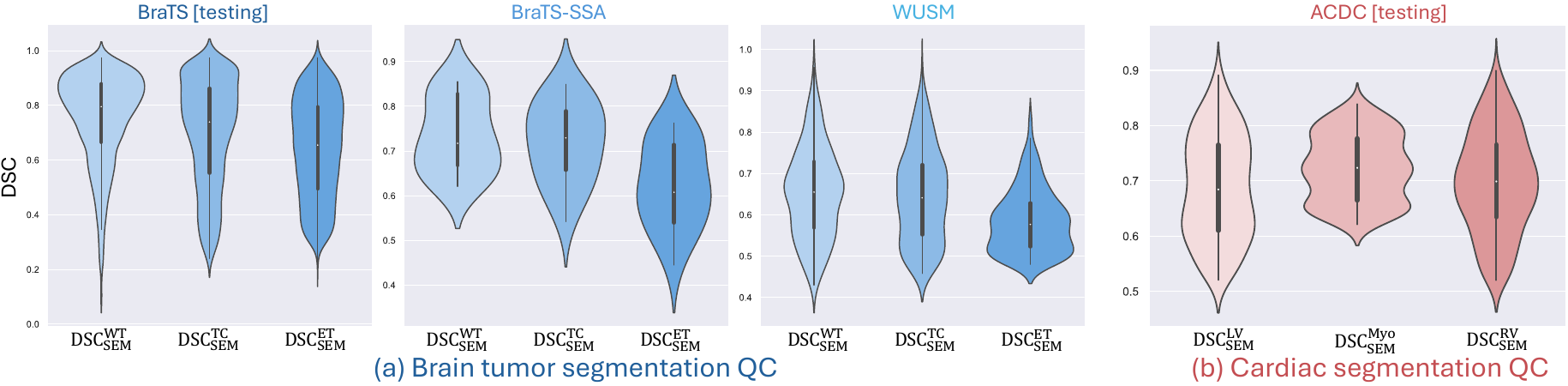}
    \caption{The distribution of the DSC (DSC$_{\text{SEM}}$) between the predicted segmentation error map and the corresponding ground truth for (a) brain tumor segmentation QC task and (b) cardiac segmentation QC task. The proposed QCResUNet accurately localized segmentation errors in terms of DSC$_{\text{SEM}}$ for all tissue classes across all datasets (refer to~\Cref{sec:err} for detailed discussion), demonstrating good generalization.}
    \label{fig:DSC_SEM_hist}
\end{figure*}

The proposed method demonstrated a strong performance gain over the state-of-the-art RCA and UE-based QC methods (\Cref{tab:brain_tumor_subj}). The performance of the RCA and the UE-based method after hyper-parameter tuning was in line with previous works~\citep{rca,jungo2020analyzing}. 
In the internal BraTS testing set, the proposed method improved the average Pearson r of NSD predictions by 48.5\% and DSC predictions by 22.1\% compared to RCA and UE-based QC, respectively.
A similar trend was observed on the external datasets, with the proposed method improving Pearson $r$ by an average of $50.9\%$ and $48.2\%$ compared to RCA and UE-based QC, respectively. Moreover, the proposed method achieved a significant reduction in the average MAE of predicting NSD and DSC compared to the RCA and UE-based methods for all results viz. internal testing results ($0.292$, $0.126$ vs. $0.064$), BraTS-SSA results ($0.281$, $0.157$ vs. $0.059$), and WUSM results ($0.275$, $0.179$ vs. $0.081$). 


\subsubsection{Cardiac MRI segmentation QC task}
Similar to the brain tumor segmentation QC task, the proposed method achieved a good performance on the cardiac segmentation QC task (see~\Cref{tab:cardiac}). Specifically, the proposed method achieved average MAEs of 0.070 and 0.057 for NSD and DSC predictions, respectively. The predicted NSD and DSC demonstrated a strong correlation with the corresponding ground truth, with Pearson r values of 0.914 and 0.955, respectively. The proposed QCResUNet also outperformed all baseline methods in the cardiac segmentation QC task. QCResUNet improved the second-best regression-based method(i.e., ResNet50~\citep{robinson2018real}) by 2.6\% and 1.8\% in Pearson r for NSD and DSC predictions, respectively. In contrast to poor performance in the brain tumor segmentation QC task, the multi-dimensional regression-based method~\citep{fournel2021medical} performed reasonably well. Despite showing inferior performance compared to other regression-based methods, it outperformed RCA and UE-based methods (see~\Cref{tab:cardiac}). 

In addition, QCResUNet significantly outperformed RCA and UE-based QC methods by an average of 10.2\% and 10.9\%. Lastly, the proposed method achieved a significant reduction in MAEs compared to all baselines (\Cref{tab:cardiac}), aligning well with the ground truth segmentation quality measures (see~\Cref{fig:scatter_DSC} and \Cref{fig:scatter_NSD}).

\begin{figure*}[!t]
    \centering
    \includegraphics[width=1.0\textwidth]{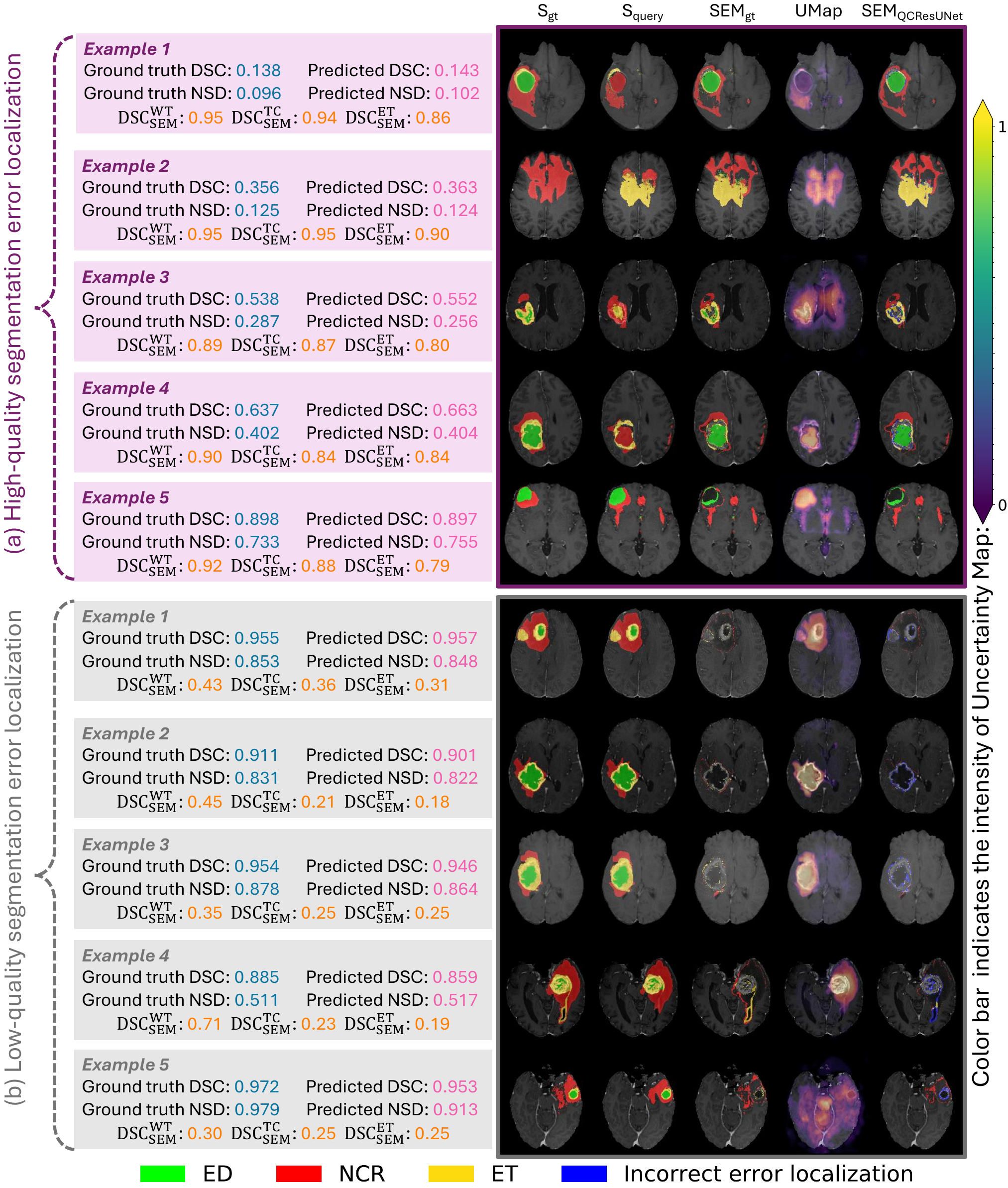}
    \caption{Examples showcasing the performance of the proposed method versus baseline methods on the brain tumor segmentation QC task: (a) high-quality segmentation error localization and (b) low-quality segmentation error localization. The color bar in the figure indicates the intensity of the uncertainty map (UMap).
    We observed that the proposed method showed better segmentation error localization than the uncertainty map. The error localization was better when dealing with low-quality query segmentation in contrast to higher-quality ones. This may be attributed to the fact that detecting few errors at the boundaries of high-quality segmentations is challenging. We kindly direct the readers to~\ref{appendix:vis} for details on how the visualization in this figure was obtained.} 
    \label{fig:vis}
\end{figure*}

\begin{figure*}[!t]
    \centering
    \includegraphics[width=0.8\textwidth]{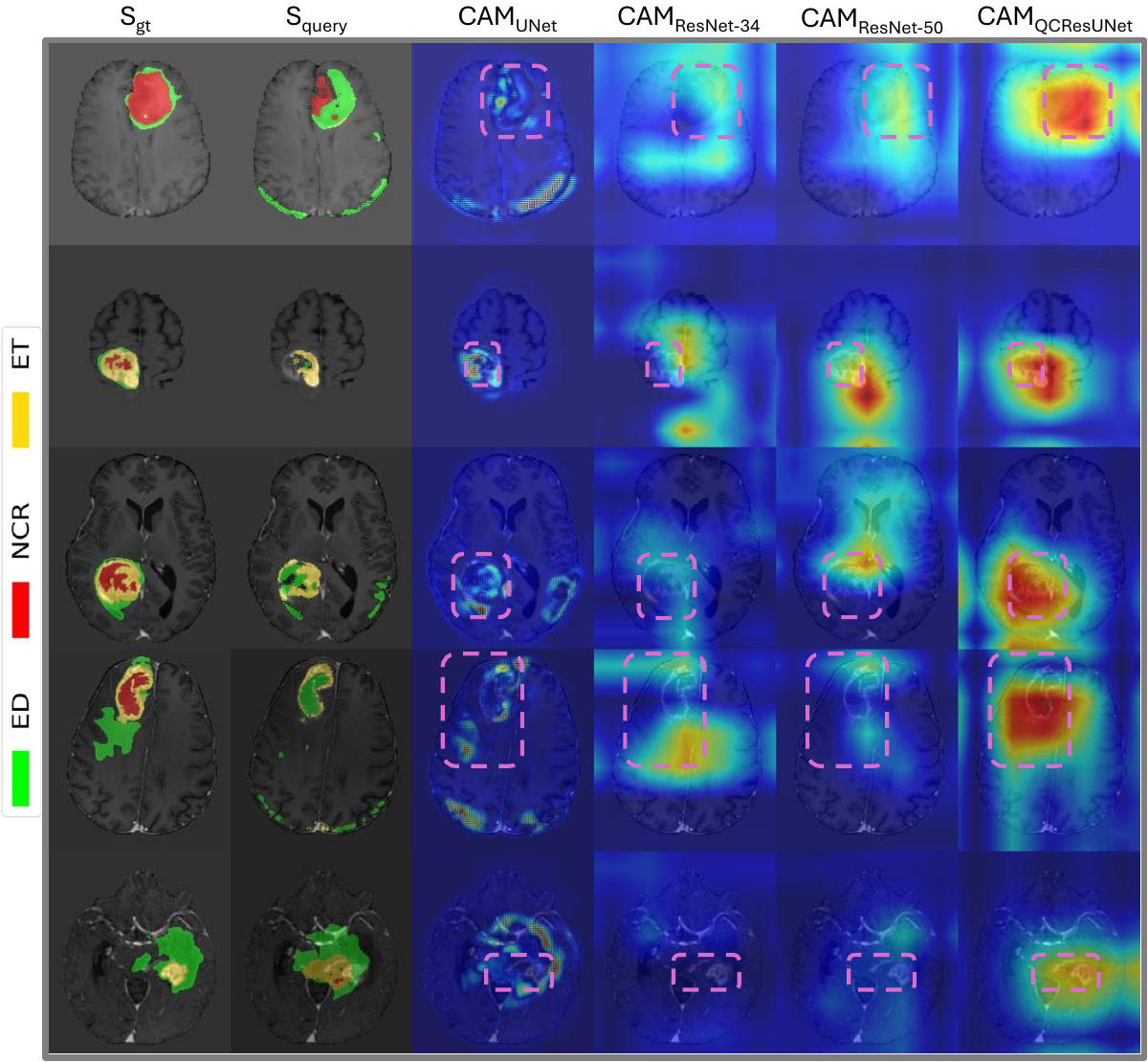}
    \caption{Examples of the class activation maps produced by Grad-CAM for different methods, with regions of interest (ROIs) highlighted in \textcolor{mypurple}{purple boxes}. The CAMs of UNet, ResNet-34, and ResNet-50 were generated based on the last convolutional feature maps. The CAMs of the proposed QCResUNet were generated from the last convolutional feature map of the ResNet encoder. We observed that the proposed QCResUNet demonstrated improved localization performance compared to the other baselines in terms of the CAM. This may be attributed to the multi-task learning framework of the proposed method (see detailed discussion in \Cref{sec:explain}). 
 }
    \label{fig:cam}
\end{figure*}

\subsection{Evaluation of segmentation error localization}
\label{sec:err}

The proposed QCResUNet achieved good voxel-level segmentation error localization in terms of DSC$_\text{SEM}$ for both brain tumor and cardiac segmentation QC tasks (see~\Cref{tab:brain_tumor_voxel}). Specifically, QCResUNet achieved an average DSC$_\text{SEM}$ of 0.769 on the BraTS internal testing set, 0.702 on the BraTS-SSA dataset, and 0.684 on the WUSM dataset in the brain tumor segmentation QC task. Despite a slight performance drop on external datasets, the results remain high quality. Inter-rater agreement studies in glioma segmentation~\citep{visser2019inter} indicate that an overlap measure above 0.7 signifies a good segmentation result. Our out-of-distribution DSC$_\text{SEM}$ results were on average around 0.7, despite the increased difficulty of the task.
In addition, QCResUNet achieved an average DSC$_{\text{SEM}}$ of 0.703 in the cardiac segmentation QC task. The detailed distribution of tissue-level DSC$_{\text{SEM}}$ (i.e., [DSC$_{\text{SEM}}^{\text{WT}}$, DSC$_{\text{SEM}}^{\text{TC}}$, DSC$_{\text{SEM}}^{\text{ET}}$] in brain tumor segmentation and [DSC$_{\text{SEM}}^{\text{LV}}$, DSC$_{\text{SEM}}^{\text{Myo}}$, DSC$_{\text{SEM}}^{\text{RV}}$] in cardiac segmentation) is shown in~\Cref{fig:DSC_SEM_hist}.

Compared to the UE-based QC method, QCResUNet improved error localization significantly by an average of 127.5\% in terms of average DSC$_{\text{SEM}}$ on the brain tumor segmentation QC task (see~\Cref{tab:brain_tumor_voxel}). Similarly, QCResUNet significantly outperformed the UE-based QC method by 52.8\% in the cardiac segmentation QC task for segmentation error localization (\Cref{tab:brain_tumor_voxel}). 

\noindent \textbf{Visual demonstration.}
Overall, the proposed QCResUNet achieved reliable localization of tissue-specific segmentation failures at the voxel level for different levels of segmentation quality (see \Cref{fig:vis}(a)). This is a unique feature of the proposed approach as none of the regression-based baseline methods can offer error localization. However, we observed that the performance of segmentation error localization might drop when the query segmentation is of high quality (\Cref{fig:vis}(b)). 
This performance drop likely arises from the inherent challenge of detecting small errors at the boundaries of high-quality segmentations. However, these cases are less critical to correct, as the query segmentation has already achieved good quality. Therefore, this drop in performance has a minor impact on the potential clinical applicability of the proposed approach. 

\subsection{Ablation analysis}
We performed ablation studies to validate the effectiveness of the proposed multi-task learning strategy and attention-based SEM aggregation mechanism. Although sharing the same network structure for subject-level segmentation quality prediction, QCResUNet outperformed ResNet-34 in both brain tumor and cardiac segmentation tasks (\Cref{tab:brain_tumor_subj} and \Cref{tab:cardiac}). In addition, we found that the performance of voxel-level segmentation error prediction deteriorated when removing the subject-level prediction task (see~\Cref{tab:brain_tumor_voxel}). This demonstrates the effectiveness of the proposed multi-task learning framework for segmentation QC tasks. 

In the voxel-level segmentation error localization task, incorporating the proposed attention-based SEM aggregation mechanism improved performance by an average of 13.9\% for the brain tumor segmentation QC task and 14.7\% for the cardiac segmentation QC task. We conjectured the improvement introduced by the proposed attention-based SEM aggregation could be attributed to the fact that the features in the last layer of the decoder contribute differently to the prediction of each tissue class error segmentation mask. 

\subsection{Explainability  analysis}
\label{sec:explain}
We hypothesized that the performance gain of the proposed QCResUnet compared to the baselines of UNet, ResNet-34, and ResNet-50 is due to joint training under both subject-level and voxel-level supervision. One potential explanation could be that the joint training allowed the proposed QCResUNet to effectively localize on segmentation errors, as suggested by the attention maps produced by Gradient-weighted Class Activation Mapping~\citep[Grad-CAM;][]{selvaraju2017grad} (see \Cref{fig:cam}). We observed that the CAM obtained from ResNet-34 and ResNet-50 did not focus on the areas of segmentation errors. Although ResNet-50 had more parameters and depth, it did not demonstrate significant improvement over ResNet-34, indicating that these factors did not improve the QC performance. 
The CAM from UNet is more sparse than that of our model and often highlights regions far away from the tumor where false positive segmentation errors occur. In contrast, the CAM produced by our model consistently highlighted tumorous areas, where significant tissue misclassifications are present.  Nonetheless, the CAM produced by our model was more spatially dispersed and blob-like. The difference in spatial support between the UNet CAM and our model’s CAM can be explained by the fact that for our QCResUNet the CAM from the encoder was used, while the CAM from the last layer was used for UNet. 



\section{Discussion and conclusions}
In this work, we proposed QCResUNet, a novel 3D CNN architecture designed for automated QC of multi-class tissue segmentation in MRI scans. To the best of our knowledge, this is the first study to provide reliable simultaneous subject-level segmentation quality predictions and voxel-level identification of segmentation errors for different tissue classes. The results suggest that the proposed method is a promising approach for 
large-scale automated segmentation QC and for guiding clinicians' feedback for refining segmentation results.

A key feature of the proposed method is the multi-task objective.  This enabled the proposed method to focus on regions where errors have occurred, leading to improved performance. This is supported by the following observations. First, we observed that the CAM of the QCResUNet encoder focused more on the regions where the segmentation error occurred compared to ResNet-34 and ResNet-50 (refer to \Cref{fig:cam}). Second, we observed that the supervision from the segmentation error prediction task can in turn guide the DSC and NSD prediction task to prioritize these error-prone regions. As suggested by the CAM, ResNet-34 and ResNet-50 achieved more accurate DSC and NSD prediction than the UNet, while the UNet performed better in segmentation error localization. The possible reason behind this is that the final average pooling layer in the UNet treats each element in the last feature map equally while ignoring the actual size of tumors. In contrast, the average pooling in the embedded feature space in ResNet-based methods operates on the abstracted quality feature maps to preserve information, which resulted in better predictive performance. 
In contrast, the joint optimization of both subject-level and voxel-level predictions by the proposed QCResUNet allows it to combine the advantages of both ResNet-based models and UNet. As a consequence, QCResUNet can simultaneously localize segmentation errors and assess the overall quality of the segmentation.

Importantly, the proposed method exhibited high generalizability when applied to unseen data, surpassing other state-of-the-art segmentation QC methods.
This was particularly true for the RCA and UE-based methods in the brain tumor segmentation QC task, which exhibited poor performance when assessing the quality of segmentation results obtained using segmentation methods different than the ones used to generate training data.
The poor generalizability of the RCA method was mainly due to the inherent difficulty of obtaining a representative reference dataset for brain tumor segmentation, which is subject to significant variability. Such significant variability may violate the underlying assumption of the RCA method that there is at least one sample in the reference dataset that can be successfully segmented given a query image-segmentation pair~\citep{rca,valindria2017reverse}, which is only valid when dealing with healthy anatomies. In the case of the cardiac segmentation QC task, which involves healthy anatomy, RCA methods achieved better performance compared to the brain tumor case when implemented with the atlas-based segmentation method.
Similar to RCA, the UE-based QC did not perform well on subject-level quality prediction as well as localizing segmentation error in the brain tumor case. This may be attributed to the fact that there is inherent variability in uncertainty maps produced by various segmentation methods on datasets with different image quality, tumor characteristics, etc. While in the cardiac segmentation QC task, which involves healthy anatomy and less variability, the UE-based showed better performance.
Additionally,
as consistent with findings in~\cite{jungo2020analyzing}, we found that the UE-based method offers limited segmentation error localization. This limitation further hinders its ability to generalize effectively. Furthermore, the UE-based method can only be used to assess segmentations obtained from deep learning models. 
Unless the deep learning segmentation method directly outputs an estimate of voxel-wise uncertainty, test time estimation of uncertainty (e.g., using MCDropout~\citep{gal2016dropout}) requires access to the model architecture and weights, which may not be possible for models deployed in clinical settings.

\noindent \textbf{Limitations}: The proposed work is not without limitations. First, though the proposed method has better generalizability than other segmentation QC methods, it may be affected by domain shift issues, which are common to all deep learning methods. As a consequence, translating the proposed to clinical practice will require monitoring of each performance to ensure reliable results. In addition, techniques that can enhance the robustness and generalizability of the proposed method to domain shifts~\citep{ganin2016domain,Carlucci_2019_CVPR} are worth further investigating in future work.

Second, a key requirement of the proposed method when applied to the brain tumor segmentation case only is the availability of all four modalities. 
The absence of certain modalities can cause a dramatic drop in QC performance (see more details illustrated in~\ref{appendix:miss}).
However, these four modalities constitute the standard of care for brain tumor diagnosis and monitoring due to the fact that they provide essential and complementary information~\citep{brats3,brats1}. Therefore, the requirement does not pose a significant practical limitation in most clinical settings. However, the problem of missing modalities can be addressed in several ways. A straightforward solution could be achieved by training models that take as input different sets of imaging modalities (i.e., have a separate model for each combination of input modalities). However, this is not a scalable solution, requiring $2^M – 1$ different networks for $M$ different modalities. Alternative approaches exist that handle missing modalities within a single model~\citep[see e.g.,][]{dorent2019hetero,shen2019brain,wang2023multi,qiu2025multimodal}. However, these require separate sets of encoders and decoders for each input modality, resulting in a computational load at least quadruple that of QCResUNet. Although handling missing modalities would enhance the applicability of the proposed method across diverse clinical settings, it is a non-trivial problem that warrants independent investigation that is beyond the scope of this paper.

Third, the proposed method was only validated on segmentation tasks involving a single object. Although we have shown that the proposed method generalized well across different segmentation tasks that include multiple tissues, more experiments are needed to evaluate how this method performs in the presence of multiple objects (e.g., multiple lesions).
Lastly, this work focused on predicting DSC and NSD as metrics to appropriately summarize segmentation quality. However, it is important to recognize that these metrics have limitations and may not be suitable for all applications. While we demonstrated the proposed method's ability to handle multiple metrics, additional research will be required to tailor the method for predicting the most appropriate metric for specific tasks~\citep{maier2024metrics}.

To conclude, we developed QCResUNet for the automated brain tumor and cardiac segmentation QC. Our proposed method is able to reliably assess segmentation quality at the subject level, while at the same time accurately identifying tissue-specific segmentation errors at the voxel level. Through multi-task learning under subject-level and voxel-level supervision, we achieved strong performance in both prediction tasks. Training the network on a large-scale dataset, which comprised segmentation results from various methods and at different levels of quality, allowed the proposed method to generalize well to unseen data. A key characteristic of the proposed method is that it is agnostic to the method used to generate the segmentation. This makes it versatile for evaluating the quality of segmentation results generated by different methods. A unique characteristic of the proposed method is its ability to accurately pinpoint the location of tissue-specific segmentation errors, thus potentially facilitating the integration of human input for refining automatically generated segmentations in clinical settings. This, in turn, has the potential to enhance clinical workflows.

\bibliographystyle{model2-names.bst}\biboptions{authoryear}
\bibliography{refs}


\appendix

\renewcommand{\thefigure}{S\arabic{figure}}
\setcounter{figure}{0}
\begin{figure*}[!t]\label{fig:hyper}
    \centering
    \includegraphics[width=1.0\textwidth]{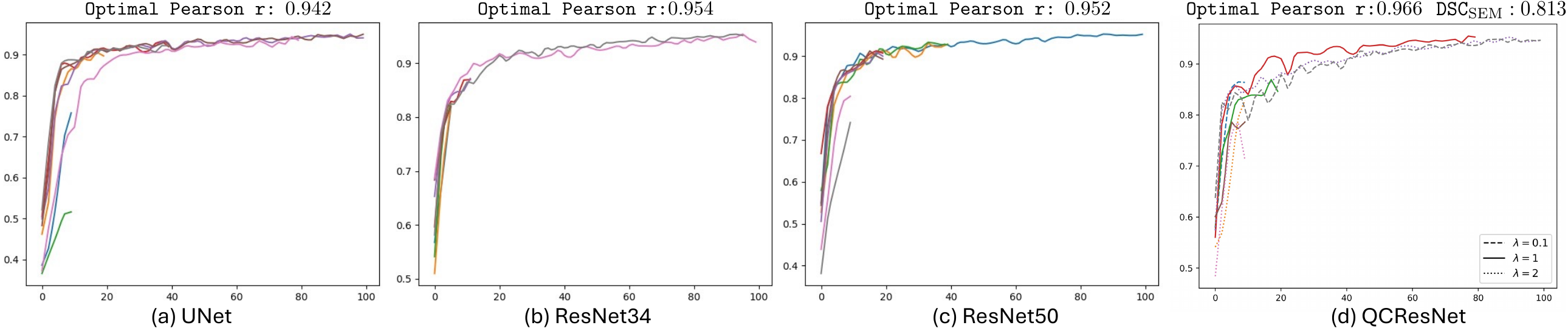}
    \caption{The learning curve of hyper-parameter tuning on the BraTS validation set, where the x-axis is the number of epochs and the y-axis is the average Pearson r for predicting DSC and NSD. Each line (differentiated by colors) in the subpanel figures indicates the Pearson r for a specific experimental trial with a particular combination of hyper-parameters.}
    \label{fig:hyper}
\end{figure*}

\section{Subject-level multi-class DSC}\label{sec:appendix_dsc}
We calculated the subject-level DSC across all tissue classes using the standard definition: 

\begin{equation}
    \text{DSC} = \frac{2 \text{TP}}{2 \text{TP} + \text{FP} + \text{FN} },
\end{equation}
where TP, FP, and FN denote true positive, false positive, and false negative, respectively.
For a certain location in the multi-class query segmentation mask ($S_{query}$), the predicted foreground label is considered a true positive if and only if it matches the corresponding ground truth label. Similarly, the predicted background label is considered a false positive if and only if it matches the true background. Otherwise, if a background label is incorrectly predicted as a foreground label, it is counted as a false positive. 

The rationale for selecting this multi-class DSC as a subject-level quality measure is to provide a single value that summarizes the overall quality of the segmentation. This is beneficial for streamlined image-driven analysis, as it allows us to filter out low-quality segmentation cases using a single threshold.

\section{Uncertainty estimation-based QC}\label{sec:appendix_a}
The uncertainty estimation based QC was implemented using the Monte Carlo Dropout (MCDropout) approach~\citep{gal2016dropout}, given its success and popularity. MCDropout was applied to the main convolution blocks in the nnUNet and DeepMedic as well as to the self-attention block in the nnFormer with a drop rate $p=0.5$, following the protocol in~\citep{jungo2020analyzing}. After running the inference of the model $T$ times, the uncertainty map of a segmentation can be computed as
\begin{equation}
    \text{UMap}_{v, c} = \frac{1}{T} \sum_{t=1}^{T} P_{v, c}^{(t)},
\end{equation}
where $v$ and $c$ denote the voxel index and the tumor tissue types, while $P_{v, c}^{(t)}$ denotes the probability map obtained from the softmax layer of a trained segmentation model. $T$ was set empirically to 50 to strike a balance between the number of Monte-Carlo samples we draw and computational complexity.

\subsection{Uncertainty-Error Overlap}\label{appendix:UE}
\label{sec:ue}
We used the uncertainty-error (UE) overlap proposed by~\cite{jungo2020analyzing} to calibrate the uncertainty map.
The UE overlap measures the overlap between the uncertainty map (UMap) and the ground-truth segmentation error map (SEM$_{gt}$) following the DSC formula:
\begin{equation}
\nonumber
    \text{UE} = \frac{2|\text{UMap} \cap \text{SEM}_{gt}|}{|\text{UMap}| + |\text{SEM}_{gt}|}, 
\end{equation}
where $|\cdot|$ denotes the cardinality of the set of voxels of each binary map. To compute the $\text{UE}$ overlap, the uncertainty map $\text{UMap}$ needs to be thresholded into a binary mask. Following the protocol in~\cite{jungo2020analyzing}, the optimal threshold value was determined by examining different thresholds in the validation set in increments of 0.05, ranging from 0.05 to 0.95. The optimal threshold values for different datasets are shown in~\Cref{tab:optimal_thres}.

\setcounter{table}{0}
\begin{table}[h]
    \centering
    \caption{The optimal threshold values used for calibrating the uncertainty map across different datasets. The optimal threshold values for brain tumor segmentation QC apply to the WT, TC, and ET tissue classes, while in the cardiac case, the threshold values pertain to the LV, Myo, and RV tissue classes.}
     \resizebox{0.475\textwidth}{!}{
    \begin{tabular}{c|ccc|c}
    \toprule
           \multicolumn{1}{c|}{} & \multicolumn{3}{c|}{\textcolor{glioma}{Brain tumor segmentation QC}} & \multicolumn{1}{c}{\textcolor{cardiac}{Cardiac segmentation QC}} \\
     \cmidrule(r){2-4} \cmidrule(r){5-5} 
     Threshold & \textcolor{brats}{BraTS [testing]}  & \textcolor{brats_ssa}{BraTS-SSA} & \textcolor{washu}{WUSM} & \textcolor{acdc}{ACDC [testing]} \\
    \toprule 
    nnUNet & [0.10, 0.10, 0.10] & [0.10, 0.10, 0.15] & [0.15, 0.10, 0.15] & [0.10, 0.10, 0.10]\\
    nnFormer & [0.10, 0.10, 0.30] & [0.30, 0.35, 0.35] & [0.40, 0.30, 0.40] & [0.10, 0.15, 0.15]\\
    DeepMedic & [0.05, 0.05, 0.05] & [0.25, 0.25, 0.20] & [0.05, 0.05, 0.10] & - \\
    \bottomrule
    \end{tabular}}
    \label{tab:optimal_thres}
\end{table}
  
\subsection{Subject-level DSC prediction}\label{appendix:subj}
Following~\cite{jungo2020analyzing}, the voxel-level uncertainty map was aggregated to subject-level Dice Similarity Coefficient (DSC) and Normalized Surface Dice (NSD) predictions by using a two-step approach. First, 102 radiomics features were automatically extracted from the uncertainty map using the \texttt{PyRadiomics}\footnote[1]{\url{https://pyradiomics.readthedocs.io}} package. Second, we trained a random forest regressor using \texttt{scikit-learn}\footnote[2]{\url{https://scikit-learn.org/stable/}} package in the training dataset to predict the subject-level DSC and NSD.

\section{Hyper-parameter tuning results}\label{appendix:hyper_param}
The hyperparameter tuning was carried out using the \texttt{Raytune}\footnote[3]{\url{https://docs.ray.io/en/latest/index.html}} package. Specifically, we utilized a Bandit-based approach for efficient resource allocations that allowed us to dedicate more resources to the more promising hyperparameter combinations~\citep{li2018hyperband,li2020system}. The learning curve for \texttt{Raytune} hyperparameter tuning in the BraTS validation set is shown in~\Cref{fig:hyper}. The optimal hyperparameter combinations determined by \texttt{Raytune} can be found in~\Cref{tab:hyperparam}.

\section{Voxel-level multi-class segmentation error mask visualization}\label{appendix:vis}
Here, we provide the details on how to obtain the visualization in~\Cref{fig:vis}. The voxel-level SEM prediction from the proposed QCResUNet is a set of binary masks (e.g., in the brain tumor segmentation task, these masks delineate the whole tumor (WT), tumor core (TC), and enhancing tumor (ET), respectively). For the purpose of visualization, we combined the binary WT, TC, and ET masks into a multi-class mask consisting of ET, NCR and ED classes. Specifically, the ED class was constructed by those voxels whose values were one in the WT binary mask and zero in the TC binary mask. The NCR class was constructed by those voxels whose values were one in the TC binary mask and zero in the ET binary mask. The ET class was constructed by those voxels whose values were one in the ET binary mask.

\section{QC performance with missing modalities}\label{appendix:miss}
\begin{table*}[!t]
\centering
\caption{Subject-level and voxel-level QC performance of the proposed QCResUNet under missing-modality settings. We reported the subject-level Pearson correlation coefficient between the predicted ground-truth NSD and DSC, denoted as $r[\text{NSD}]$ and $r[\text{DSC}]$, respectively. For voxel-level QC performance, we reported the average of DSC$_{\text{SEM}}^{\text{WT}}$, DSC$_{\text{SEM}}^{\text{TC}}$, and DSC$_{\text{SEM}}^{\text{ET}}$, denoted as DSC$_{\text{SEM}}^{\text{avg}}$. \textbullet~and \textopenbullet~denote available and missing modalities, respectively.}
\resizebox{0.95\textwidth}{!}{
\begin{tabular}{cccc|ccc|ccc|ccc}
\toprule
\multicolumn{4}{c|}{Modalities} & \multicolumn{3}{c|}{\textcolor{brats}{BraTS-testing}} & \multicolumn{3}{c|}{\textcolor{brats_ssa}{BraTS-SSA}}& \multicolumn{3}{c}{\textcolor{washu}{WUSM}}  \\
FLAIR & T1c & T1 & T2 & $r$[NSD] & $r$[DSC] & DSC$_{\text{SEM}}^{\text{avg}}$ & $r$[NSD] & $r$[DSC] & DSC$_{\text{SEM}}^{\text{avg}}$ & $r$[NSD] & $r$[DSC] & DSC$_{\text{SEM}}^{\text{avg}}$  \\
\midrule
\textbullet & \textopenbullet & \textopenbullet & \textopenbullet & 0.787 & 0.811 & 0.499 & 0.776 & 0.760 & 0.387 & 0.709 & 0.706 & 0.338 \\
\textopenbullet & \textbullet & \textopenbullet & \textopenbullet & 0.654 & 0.618 & 0.515 & 0.597 & 0.583 & 0.409 & 0.593 & 0.626 & 0.371 \\
\textopenbullet & \textopenbullet & \textbullet & \textopenbullet & 0.541 & 0.493 & 0.475 & 0.551 & 0.533 & 0.384 & 0.547 & 0.477 & 0.350 \\
\textopenbullet & \textopenbullet & \textopenbullet & \textbullet & 0.695 & 0.581 & 0.429 & 0.697 & 0.635 & 0.363 & 0.622 & 0.538 & 0.349 \\
\midrule
\textbullet & \textbullet & \textopenbullet & \textopenbullet & 0.887 & 0.916 & 0.669 & 0.821 & 0.856 & 0.628 & 0.827 & 0.757 & 0.600 \\
\textbullet & \textopenbullet & \textbullet & \textopenbullet & 0.875 & 0.875 & 0.546 & 0.854 & 0.836 & 0.494 & 0.797 & 0.709 & 0.444 \\
\textbullet & \textopenbullet & \textopenbullet & \textbullet & 0.885 & 0.872 & 0.536 & 0.860 & 0.809 & 0.490 & 0.826 & 0.726 & 0.447 \\
\textopenbullet & \textbullet & \textbullet & \textopenbullet & 0.617 & 0.695 & 0.588 & 0.578 & 0.603 & 0.520 & 0.613 & 0.650 & 0.480 \\
\textopenbullet & \textbullet & \textopenbullet & \textbullet & 0.703 & 0.735 & 0.599 & 0.664 & 0.653 & 0.506 & 0.668 & 0.699 & 0.479 \\
\textopenbullet & \textopenbullet & \textbullet & \textbullet & 0.604 & 0.608 & 0.448 & 0.631 & 0.606 & 0.485 & 0.593 & 0.514 & 0.449 \\
\midrule
\textbullet & \textbullet & \textbullet & \textopenbullet & 0.912 & 0.943 & 0.715 & 0.861 & 0.891 & 0.638 & 0.835 & 0.770 & 0.594 \\
\textbullet & \textbullet & \textopenbullet & \textbullet & 0.930 & 0.948 & 0.709 & 0.904 & 0.892 & 0.624 & 0.847 & 0.758 & 0.592 \\
\textbullet & \textopenbullet & \textbullet & \textbullet & 0.901 & 0.887 & 0.547 & 0.878 & 0.836 & 0.497 & 0.822 & 0.712 & 0.446 \\
\textopenbullet & \textbullet & \textbullet & \textbullet & 0.776 & 0.789 & 0.629 & 0.639 & 0.708 & 0.515 & 0.622 & 0.719 & 0.489 \\
\midrule
 & Average & & & 0.769 &  0.769 & 0.565 & 0.736 & 0.728 & 0.496 & 0.709 & 0.669 & 0.459 \\
\bottomrule
\end{tabular}}
\label{tab:miss}
\end{table*}
We reported the subject-level and voxel-level QC performance under the missing-modality settings in~\Cref{tab:miss}. Compared to the model that takes as input all modalities (as reported in~\Cref{tab:brain_tumor_subj}), the subject-level QC performance on the BraTS internal testing dataset decreased by an average of 19.6\% and 20.56\% in terms of Pearson correlation coefficients for NSD and DSC prediction, respectively, across all modality combinations. Similarly, on the BraTS-SSA dataset, the Pearson correlation coefficients for NSD and DSC dropped by an average of 22.9\% and 22.9\%, and by 24.6\% and 26.6\% on the WUSM dataset. A similar trend was observed for the voxel-level QC performance: the DSC for SEM across all sub-tissue types dropped by an average of 26.5\%, 29.3\%, and 32.9\% on the BraTS testing, BraTS-SSA, and WUSM datasets, respectively.

Although a significant performance drop was observed in the setting of missing modalities, our QCResUNet showed a reasonable performance when at least two modalities are present, achieving an average subject-level Pearson correlation for NSD and DSC of 0.774 and 0.766, respectively, across all datasets. Likewise, the proposed method can achieve an average voxel-level QC performance of 0.547 in terms of DSC for SEM. However, if only one modality is present, especially when T1c or FLAIR is missing, the proposed method showed a relatively poor performance, with an average subject-level Pearson correlation coefficient of 0.609 and 0.543 for NSD and DSC. This is because the T1 and T2 modalities do not provide sufficient information for delineating the enhancing tumor and edema. 

\section{Computational Complexity}\label{appendix:computation}
\begin{table}[!t]
     \caption{Benchmark of computational complexity of different models.}
    \centering
    \begin{tabular}{l|c|c}
        \toprule
         Model & Number of Parameters (M) & FLOPs (G) \\
         \midrule
         ResNet-34 & 8.36 & 249.64 \\
         ResNet-50 & 15.94 & 302.86 \\
         UNet & 16.60 & 670.24\\
         QCResUNet & 16.59 & 296.64 \\
         \bottomrule
    \end{tabular}
   
    \label{tab:computation}
\end{table}
We provided a benchmark of computational complexity for the different architectures we used in this paper in~\Cref{tab:computation}. The proposed QCResUNet has a parameter and FLOPs count comparable to that of ResNet-50, but nearly doubles the number of parameters of ResNet-34 as a result of the added decoder modules. However, our QCResUNet incorporated a relatively shallow decoder with only three upsampling stages—compared to five in the UNet—resulting in a modest increase in FLOPs relative to ResNet-34, but significantly fewer FLOPs than UNet. We would also like to point out that the proposed attention mechanism is efficient in terms of parameters, as it only involves $1 \times 1 \times 1$ convolutions.


\end{document}